\documentclass[reprint,aps,pra]{revtex4-1}
\usepackage{amsmath}
\usepackage{graphicx}
\usepackage{dcolumn}
\usepackage{bm}
\usepackage{color}
\usepackage{subfig}

\begin{document}

\title{Splitting of nonlinear-Schr\"{o}dinger breathers by linear and
nonlinear localized potentials}
\author{Oleksandr V. Marchukov and Boris A. Malomed}
\affiliation{ Department of Physical Electronics, School of Electrical
Engineering, Faculty of Engineering, and Center for Light-Matter Interaction,
Tel Aviv University, 6997801 Tel Aviv,
Israel}
\author{Vladimir A. Yurovsky}
\affiliation{School of Chemistry, Tel Aviv University, 6997801 Tel Aviv,
Israel}
\author{Maxim Olshanii and Vanja Dunjko}
\affiliation{Department of Physics, University of Massachusetts Boston,
Boston, Massachusetts 02125, USA}
\author{Randall G. Hulet}
\affiliation{Department of Physics and Astronomy, Rice University,
Houston,Texas 77005, USA}

\begin{abstract}
We consider evolution of one-dimensional nonlinear-Schr\"{o}dinger (NLS)
two-soliton complexes (breathers) with narrow repulsive or attractive
potentials (barrier or well, respectively). By means of systematic
simulations, we demonstrate that the breather may either split into
constituent fundamental solitons (\textit{fragments}) moving in opposite
directions, or bounce as a whole from the barrier. A critical initial
position of the breather, which separates these scenarios, is predicted by
an analytical approximation. The narrow potential well tends to trap the
fragment with the larger amplitude, while the other one escapes. The
interaction of the breather with a nonlinear potential barrier is also
considered. The ratio of amplitudes of the emerging free solitons may be
different from the $3:1$ value suggested by the exact NLS solution,
especially in the case of the nonlinear potential barrier. Post-splitting
velocities of escaping solitons may be predicted by an approximation based
on the energy balance.
\end{abstract}

\maketitle

\preprint{APS/123-QED}

\section{\label{sec:intro}Introduction}

It is commonly known that solitons, i.e., robust self-trapped modes in
nonlinear media, are fundamental excitations, and in many cases represent
ground states, in a broad variety of physical settings \cite{KA,DP,Y}.
Additional interest to solitons has been attracted by the creation of
localized matter-wave states in Bose-Einstein condensates (BECs) of
ultracold ${}^{7}\mathrm{Li}$ \cite{strecker2002,khaykovich2002,strecker2003}
and ${}^{85}\mathrm{Rb}$ \cite{cornish2006, marchant2013} atoms. These
solitons were produced in the quasi-one-dimensional (quasi-1D) form, by
loading the atomic BEC into a strongly anisotropic (cigar-shaped) potential
trap. The latter setting is accurately approximated, in the framework of the
mean-field theory, by the Gross-Pitaevskii equations (GPEs) \cite{Pit}. In
the case of attractive interactions between atoms, the GPE, being similar to
the integrable nonlinear Schr\"{o}dinger (NLS) equation, gives rise to
well-known soliton solutions~\cite{zakharov1972, zakharov1984, satsuma1974}.
\begin{figure}[tbp]
\centering
\subfloat{\includegraphics[width=0.250\textwidth]{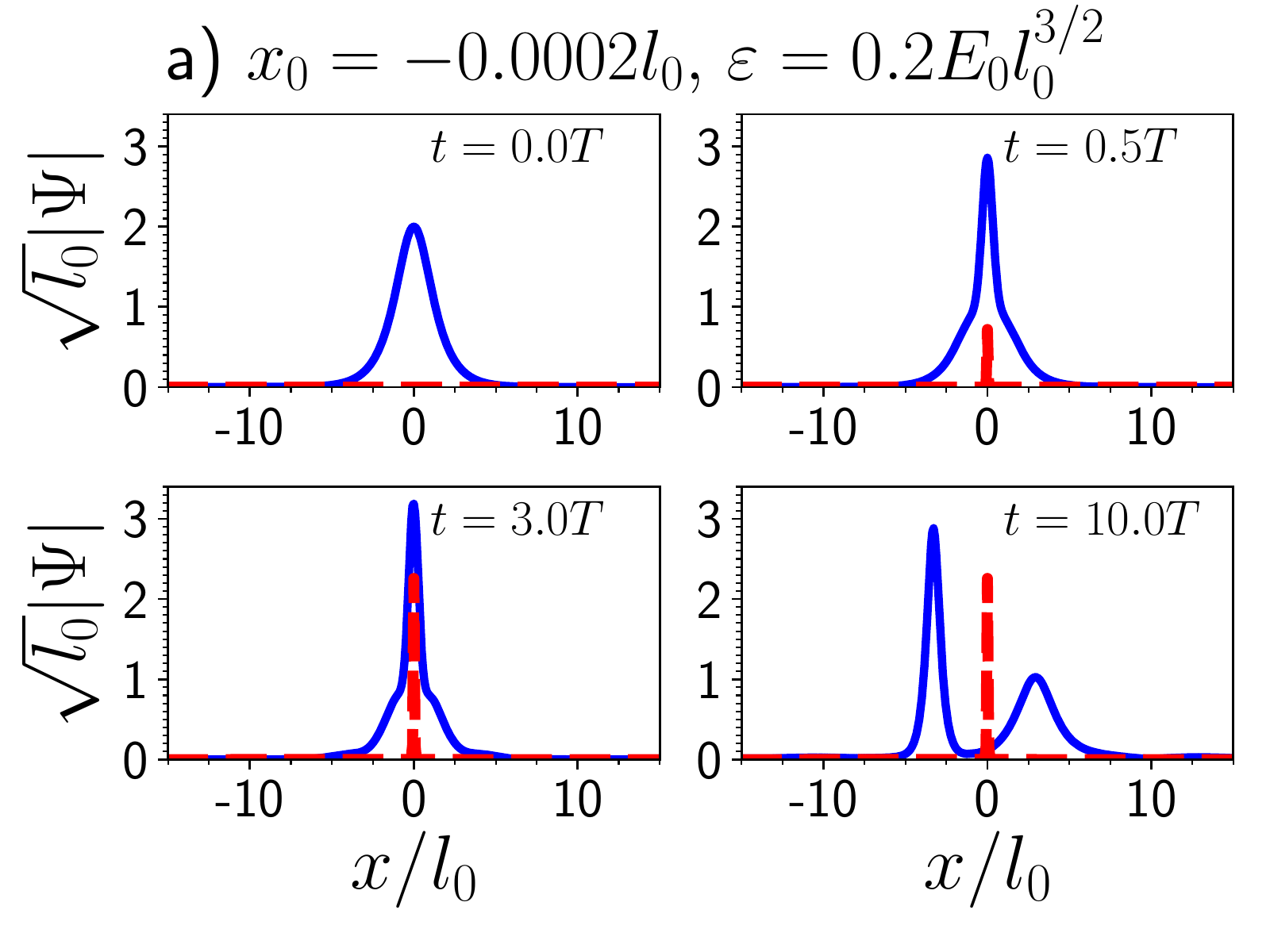}}~%
\subfloat{\includegraphics[width=0.250\textwidth]{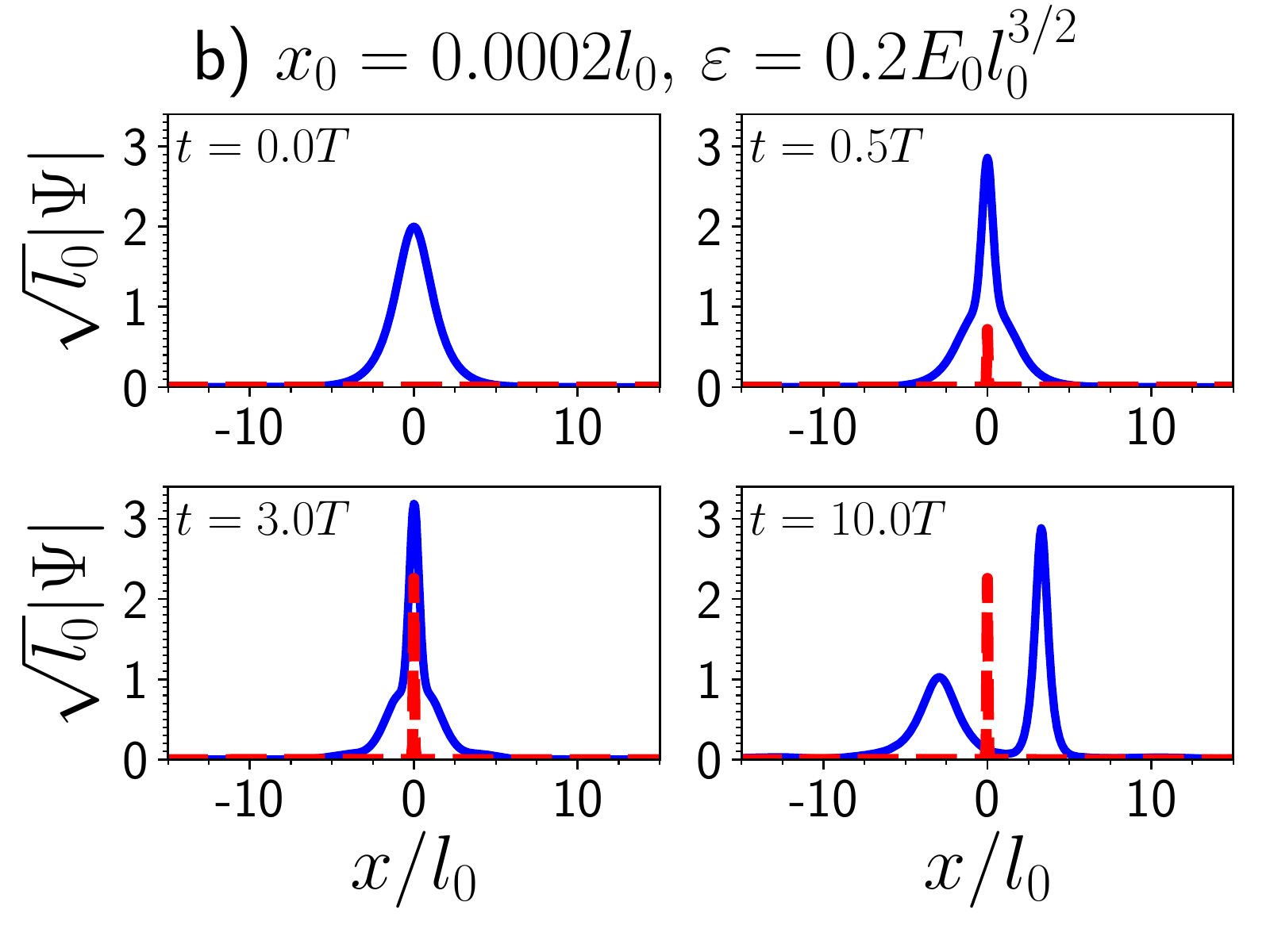}} ~\\[-1ex]
\subfloat{\includegraphics[width=0.250\textwidth]{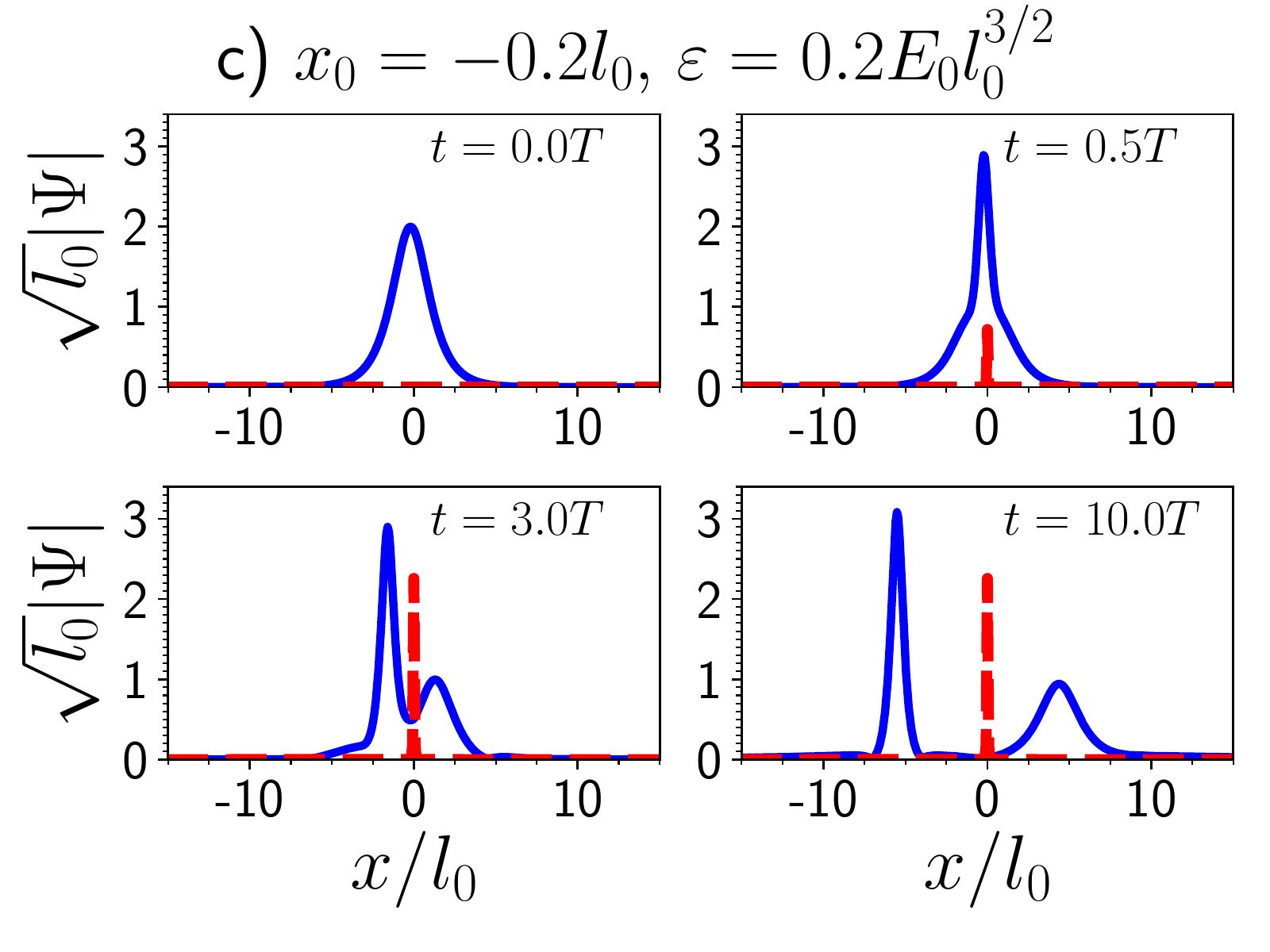}}~%
\subfloat{\includegraphics[width=0.250\textwidth]{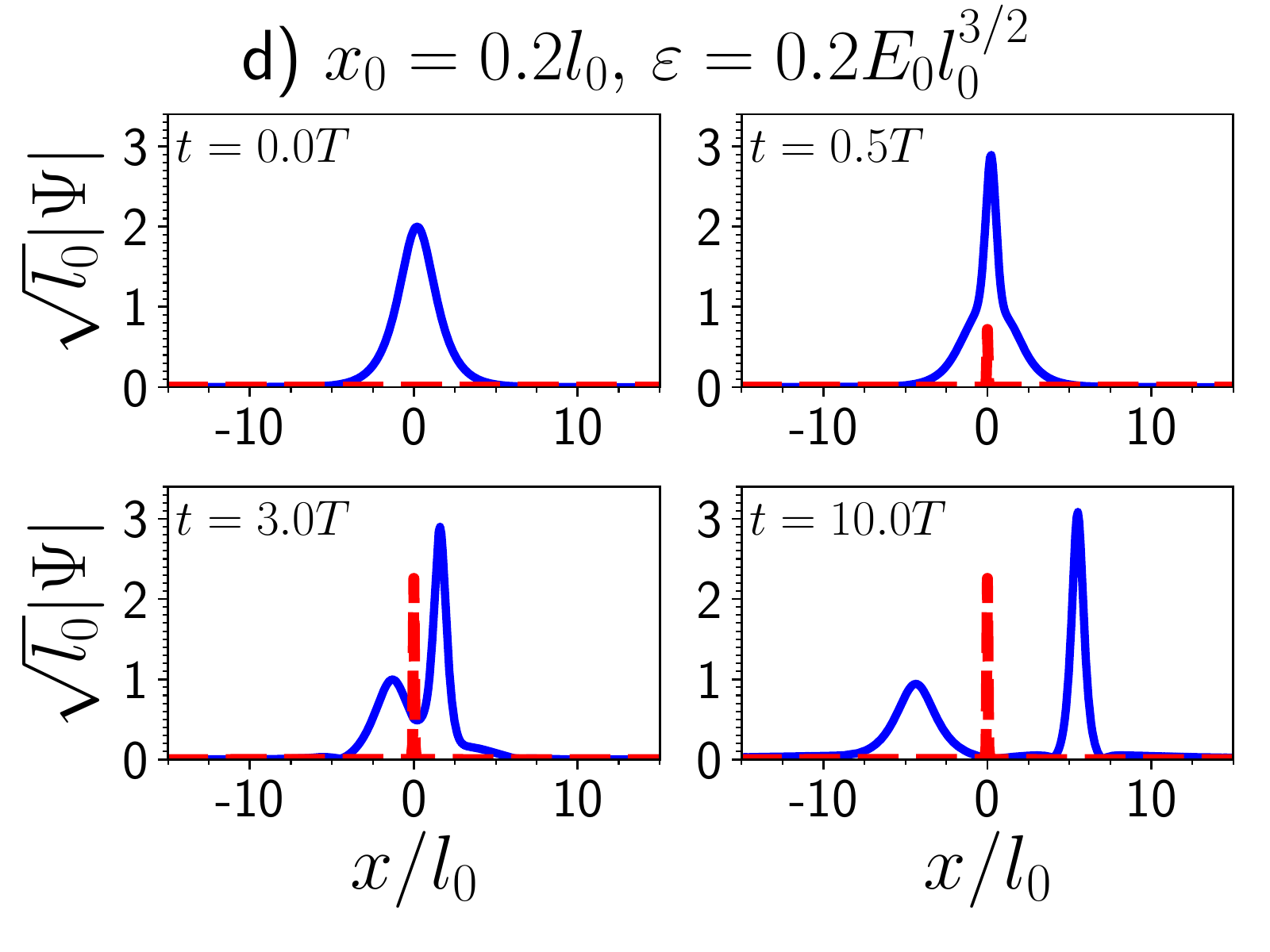}} \\[-1ex]
\subfloat{\includegraphics[width=0.250\textwidth]{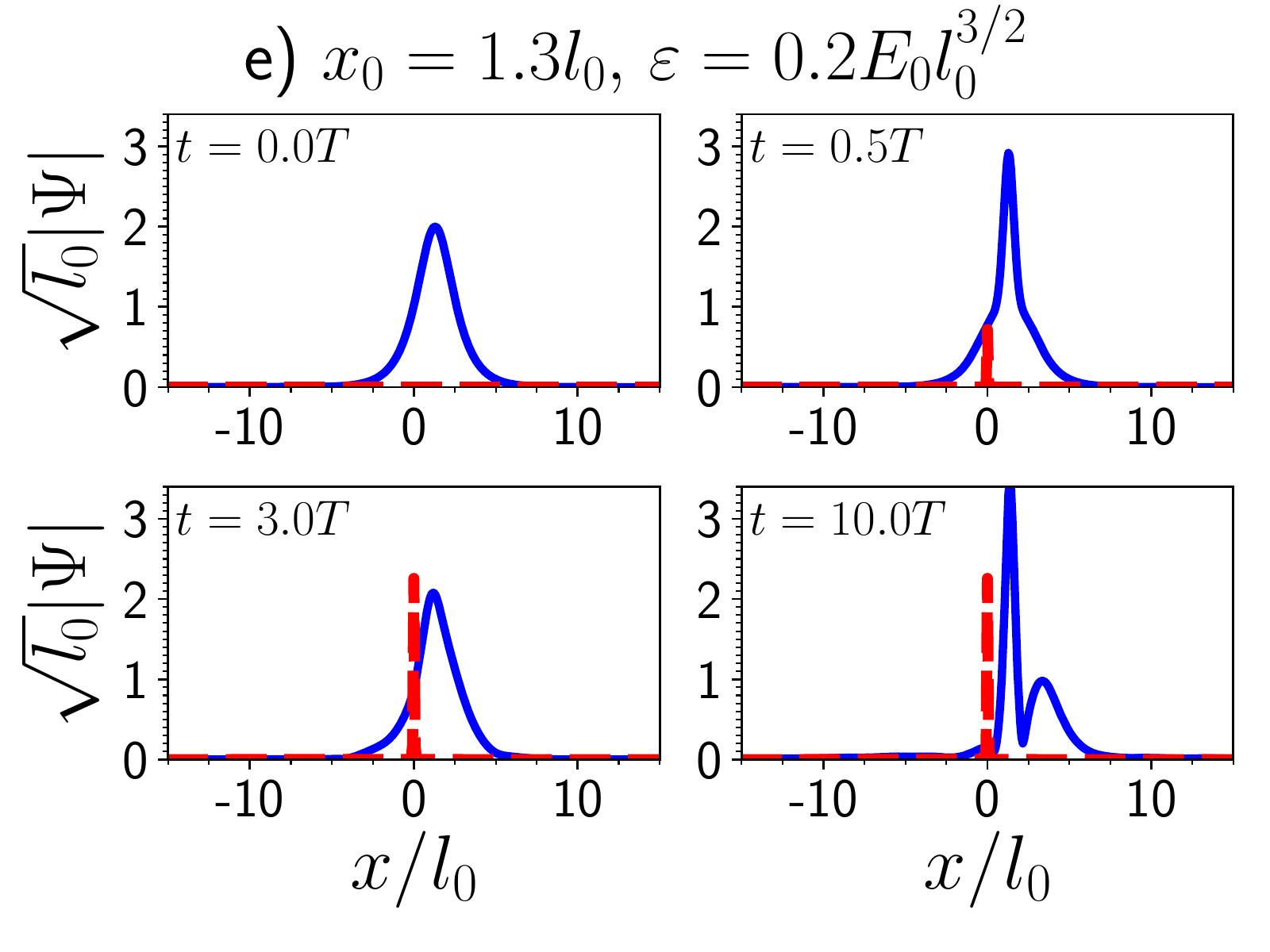}}
\caption{Snapshots of direct simulations of splitting of input (\protect\ref%
{t=0}) under the action of the the linear potential barrier with $\protect%
\varepsilon =0.2$ in Eq. (\protect\ref{nls}), and different values of shift $%
x_{0}$: a) $x_{0}=-0.0002$ b) $x_{0}=0.0002$; c) $x_{0}=-0.2$; d) $x_{0}=0.2$%
; e) $x_{0}=1.3$. The snapshots are displayed at times indicated in the
panels. The blue solid lines represent profiles of $|\Psi (x)|$ at the
respective moments of time, while the red dashed line is the shape of the
splitting barrier. Panels a) and b) demonstrate that even an extremely small
shift of the initial position of the input from the central point leads to
strong (although slowly developing) breaking of the spatial symmetry and
fission of the breather.}
\label{dir_sim}
\end{figure}

In this work we focus on two-soliton solutions, also known as \textit{%
breathers}. Using the inverse-scattering-transform~technique which applies
to many integrable equations \cite{zakharov1972, zakharov1984, satsuma1974},
one can show that, in the simplest case the breather may be considered as a
nonlinear superposition of two fundamental solitons, with the ratio of
amplitudes $3:1$ and zero relative velocity (it is relevant to mention that
the Lieb-Liniger model \cite{LLM,LLM-review}, i.e., the integrable quantum
version of the NLS equation, does not produce spatially-localized solitons
and breathers, supporting solely spatially-homogeneous strings of bound
quantum particles \cite{Volodya}). Thus, it is reasonable to expect that
perturbations, represented by additional terms in the NLS equation, which
break its integrability, may split the breather into constituent solitons (%
\textit{fragments}) \cite{HS-resonant,Khaykovich, dunjko2015, malomed2018,
we}.

In many physical settings, an important type of integrability-breaking
perturbations added to the NLS equation is represented by narrow potential
barriers \cite{barrier1}-\cite{barrier6} or wells \cite{well2,well1}. In
this work we aim to consider the interaction of two-soliton breathers with a
narrow linear barrier and well, as well as with a nonlinear barrier. This
setting directly represents experimentally relevant situations. In optics, a
light beam in the form of the spatial two-soliton can propagate along the
medium's defect in the form of the barrier or well. In a BEC, a breather
traveling in a cigar-shaped trap can hit a splitting barrier, which is a
typical situation in matter-wave interferometers \cite{interferometer1}-\cite%
{interferometer7}. The interaction with localized barriers and wells was
considered many times for fundamental solitons \cite{barrier1}-\cite{HS},
but it was not previously studied for breathers. In particular, because
splitting an incident fundamental soliton in two fragments by a barrier
underlies the operation of matter-wave interferometers, it is definitely
relevant to consider a similar effect for two-soliton breathers, for which
the splitting is a much more natural, hence also more controllable, outcome
of the interaction with the localized potential.

The breather-barrier interaction is controlled by two main parameters: the
barrier's height and its initial shift with respect to the breather's
center. By means of systematic simulations, we explore the outcome of the
interaction, in the parameter space of the model. The results make it
possible to identify two distinct regimes, \textit{viz}., the breather
splitting into a pair of constituent fundamental solitons, moving in
opposite directions, or survival of the breather, moving as a whole in a
certain direction. In the former case, amplitudes of the fragment solitons
keep the above-mentioned $3:1$ ratio only approximately, sometimes featuring
a relatively complex pattern in the distribution of their amplitudes and
velocities.

The paper is organized as follows: In Section~\ref{sec:model} we formulate
the problem, introducing the original breather and defining the linear and
nonlinear localized potentials with which the breather interacts. In Section~%
\ref{sec:methods}, we outline the numerical procedure and present an
analytical evaluation of a threshold value of the initial position of the
center of the breather with respect to the barrier, that separates the
unidirectional and bidirectional motion of the products of the
breather-barrier interaction. We also provide an estimate for the velocities
of the outgoing solitons from the energy-balance analysis. In Section~\ref%
{sec:results} we present results of systematic simulations of the
interaction for the linear and nonlinear barriers. Finally, in Section~\ref%
{sec:conclusions} we summarize findings produced by the present analysis and
discuss possibilities for future work.

\section{\label{sec:model}The model}

\textit{Linear potential barrier and well}. Starting with the formulation of
the BEC model, we consider a nearly 1D Bosonic condensate, which in the
mean-field approximation is governed by the GPE with an attractive
nonlinearity \cite{Pit}. In scaled variables, it takes the form of
\begin{equation}
i\frac{\partial \Psi }{\partial t}=-\frac{1}{2}\frac{\partial ^{2}\Psi }{%
\partial x^{2}}-g|\Psi |^{2}\Psi +\varepsilon f(t)\delta (x)\Psi ,
\label{nls}
\end{equation}%
where $\Psi (x,t)$ is the single-particle wave function, $x$ the spatial
coordinate, $t$ time, and $\varepsilon >0$ or $\varepsilon <0$ is,
respectively, the strength of the narrow potential barrier or well, which is
represented by the delta-function, to be replaced by its usual regularized
version in numerical simulations,%
\begin{equation}
\delta \left( x\right) =\frac{1}{\sqrt{\pi }a}\exp \left( -\frac{%
x^{2}}{a^{2}}\right) ,  \label{U}
\end{equation}%
with small width $a$, fixed to be $a=0.05$ in the numerical simulations
performed below.

Function $f(t)$ accounts for gradually increasing the linear potential, in
the course of a rise time, $\tau $. In this work, we adopt it in form of%
\begin{equation}
f(t)=\left\{
\begin{array}{c}
t/\tau ,~\ \mathrm{at}~\ 0<t<\tau , \\
1,~~\mathrm{at}~\ t>\tau ~.%
\end{array}%
\right.  \label{f}
\end{equation}%
In particular, in the BEC setting the local potential is applied by a
focused laser beam \cite{Cuevas}, whose power gradually increases from zero
at $t=0$ to a maximum value at $t=\tau $.

The strength of the self-attraction in Eq. (\ref{nls}), which
builds matter-wave solitons, is $g=-2\hbar a_{s}\omega _{\perp }>0$, where $%
a_{s}<0$ is the respective \textit{s}-wave scattering length, and $\omega
_{\perp }$ the transverse trapping frequency of the confinement, that
reduces the effective geometry from 3D to 1D~\cite{olshanii1998}. Note that
we consider the transverse trapping frequency to be sufficiently large for
neglecting the next-order terms in the interaction strength, such as the one induced by the external confinement \cite{olshanii1998}. In terms of the underlying physical parameters (including atomic mass $m$ and $g$), the units of length, $l_{0}$%
, time, $t_{0}$, and energy, $E_{0}$, of the model may be defined as $%
l_{0}=\hbar ^{2}/\left( mg\right) $, $t_{0}=\hbar ^{3}/\left( mg^{2}\right) $%
, and $E_{0}=mg^{2}/\hbar ^{2}$, respectively. Note that,
in such units, Eq. (\ref{nls}) can be further rescaled to set $g=1$, which
we use below.

In the above-mentioned optics model, variable $t$ in Eq. (\ref{nls})
actually represents the propagation distance, while $x$ is the transverse
coordinate in the planar waveguide, $g$ is the usual Kerr-nonlinearity
coefficient (in the scaled form), and the barrier/well represents a narrow
stripe with a lower/higher value of the local refractive index, which is
subject to longitudinal modulation as per Eq. (\ref{f}). Similarly, the
nonlinear barrier may be realized as a stripe of a defocusing material built
into a self-focusing waveguide.

The same setting with a linear barrier may also be implemented in the
temporal domain, i.e., in a nonlinear optical fiber, using co-propagation of
light signals at two different carrier wavelengths with nearly equal group
velocities $V_{\mathrm{gr}}$, but opposite signs of the group-velocity
dispersion (GVD), which can be selected on two sides of a zero-GVD point, as
suggested in another context in Ref. \cite{Arkady}. In this case, $t$ in Eq.
(\ref{nls}) is realized as the propagation distance along the fiber, while $%
x\equiv \theta -t/V_{\mathrm{gr}}$, with physical time $\theta $ \cite{KA}.
The optical breather is then launched in the channel with anomalous GVD,
while an effective barrier is induced, via the cross-phase modulation (XPM),
by a narrow dark soliton coupled into the normal-GVD channel, supported by a
high-power continuous-wave background. As concerns the potential well, it
may be realized by launching signals in two mutually orthogonal circular
polarizations at the same carrier wavelength, with anomalous GVD. In the
latter case, the breather is launched in one polarization, while the
effective well is induced, via XPM, by a high-power narrow bright soliton
carried by the other polarization.

In the case of $\varepsilon =0$, Eq.~\eqref{nls} with $g=1$ has an exact
periodic $2$-soliton solution (breather),
\begin{widetext}
\begin{equation}
\psi _{\mathrm{br}}(x,t)=4Ae^{iA^{2}t/2}\frac{\cosh \left[ 3A\left(
x-x_{0}\right) \right] +3\exp \left( 4iA^{2}t\right) \cosh \left[ A(x-x_{0})%
\right] }{\cosh \left[ 4A\left( x-x_{0}\right) \right] +4\cosh \left[
2A\left( x-x_{0}\right) \right] +3\cos \left( 4A^{2}t\right) },
\label{exact}
\end{equation}
\end{widetext}where $A$ is an arbitrary real amplitude, which also
determines the oscillation period, $T=\pi /(2A^{2})$, and $x_{0}$ is the
initial displacement of the breather's center from the point where the
narrow potential barrier (or well) is placed. In the simulations reported
below we set the rise time in Eq. (\ref{f}) to be $\tau =T$, although other
values of $\tau $ yield very similar results.

The breather given by Eq. (\ref{exact}) may be considered as a nonlinear
superposition of two fundamental solitons with amplitudes%
\begin{equation}
A_{1}=3A,~A_{2}=A,  \label{A1A2}
\end{equation}%
and zero velocities. On the other hand, solitons moving with velocities $%
c_{1,2}$ are represented by solutions
\begin{gather}
\psi _{\mathrm{sol}}\left( x,t\right) =A_{1,2}~\mathrm{sech}\left(
A_{1,2}\left( x-c_{1,2}t\right) \right) \times  \notag \\
\exp \left[ i\left( A_{1,2}^{2}-c_{1,2}^{2}\right) t/2\right]  \label{sol}
\end{gather}%
(overlap between them may be neglected provided that $\left\vert
c_{1}-c_{2}\right\vert t\gg A_{1,2}^{-1}$). Note that the integral norm of
the breather,
\begin{equation}
\int_{-\infty }^{+\infty }\left\vert \psi _{\mathrm{br}}\left( x,t\right)
\right\vert ^{2}dx=8A,  \label{Nbr}
\end{equation}%
is exactly equal to the summary norm of two solitons (\ref{sol}),
\begin{equation}
N_{\mathrm{sol}}^{(1)}+N_{\mathrm{sol}}^{(2)}\equiv 2A_{1}+2A_{2}\equiv 8A.
\label{Nsol}
\end{equation}%
At $t=0$, the breather in Eq.~\eqref{exact} takes the simplest shape,
\begin{equation}
\psi _{\mathrm{br}}(x,0)=2A~\mathrm{sech}(A(x-x_{0})),  \label{t=0}
\end{equation}
which is used as the initial condition in simulations reported below. In a
more general setting, one can use the input given by Eq. (\ref{exact}) at $%
t=t_{0}\neq 0$. Additional simulations demonstrate that variation of $t_{0}$
does not essentially modify systematic results presented below for $t_{0}=0$%
.
\begin{figure}[tbp]
\centering
\subfloat{\includegraphics[width=0.250\textwidth]{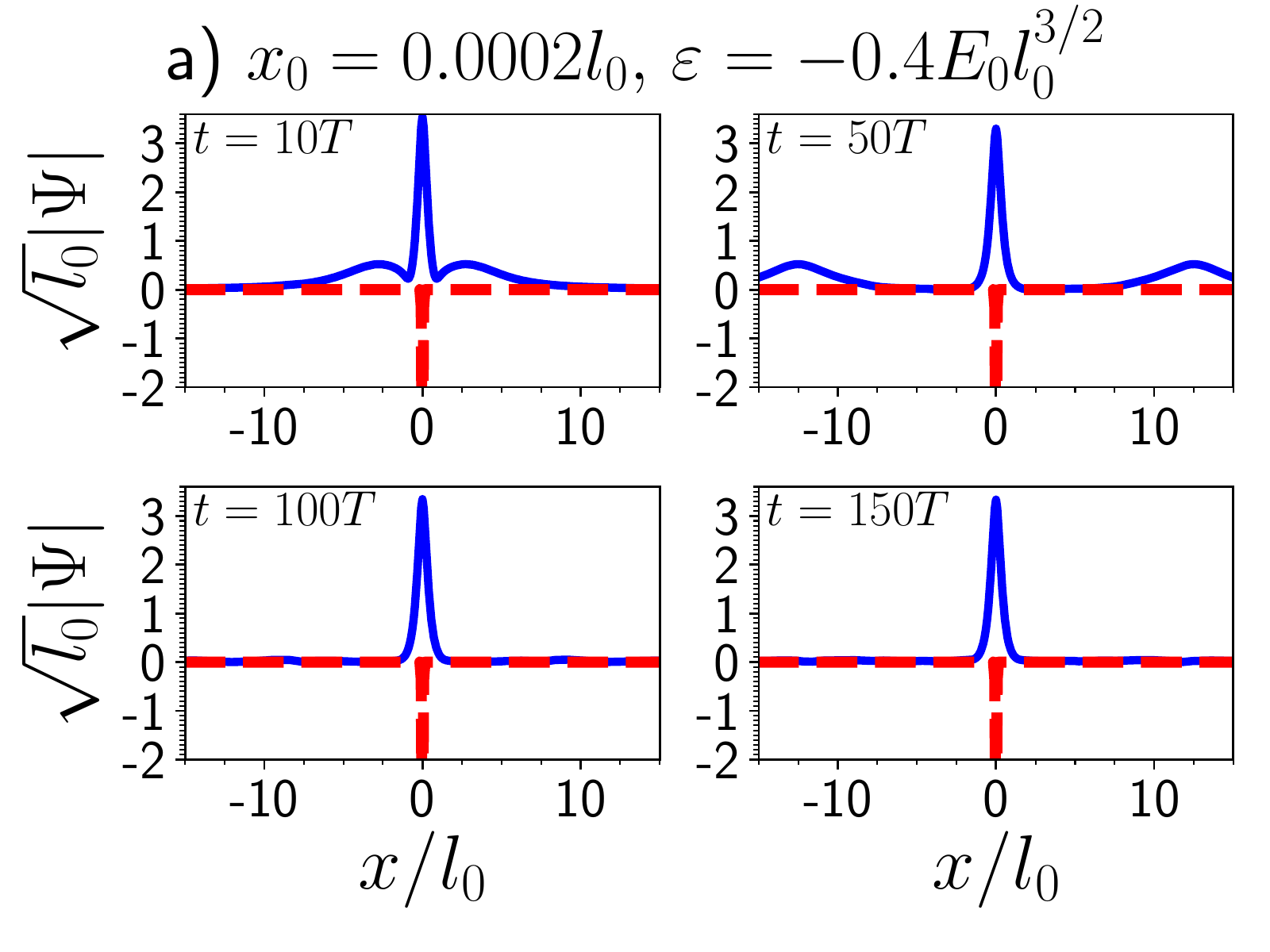}} ~%
\subfloat{\includegraphics[width=0.250\textwidth]{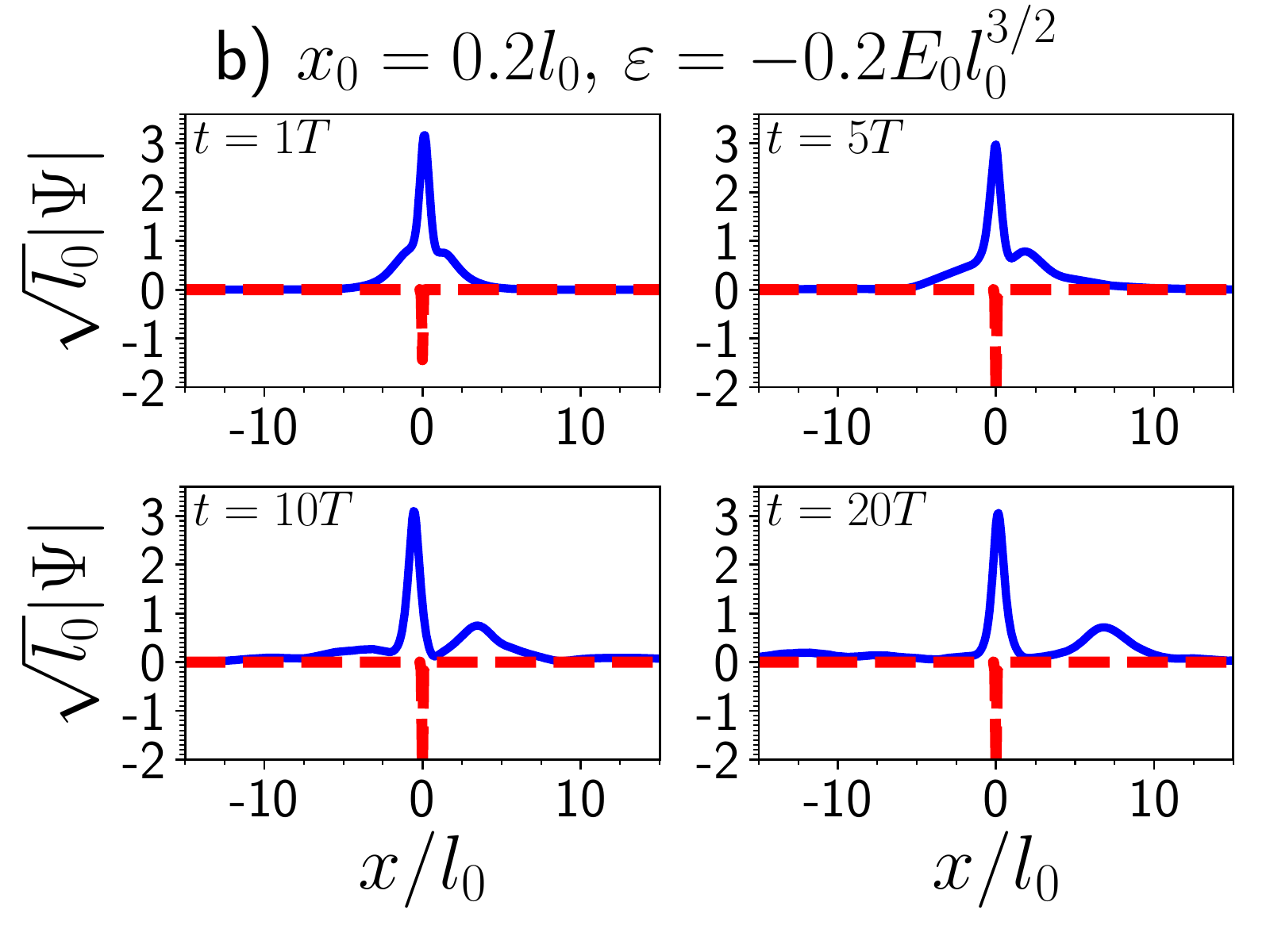}}~\\[-3ex]
\subfloat{\includegraphics[width=0.250\textwidth]{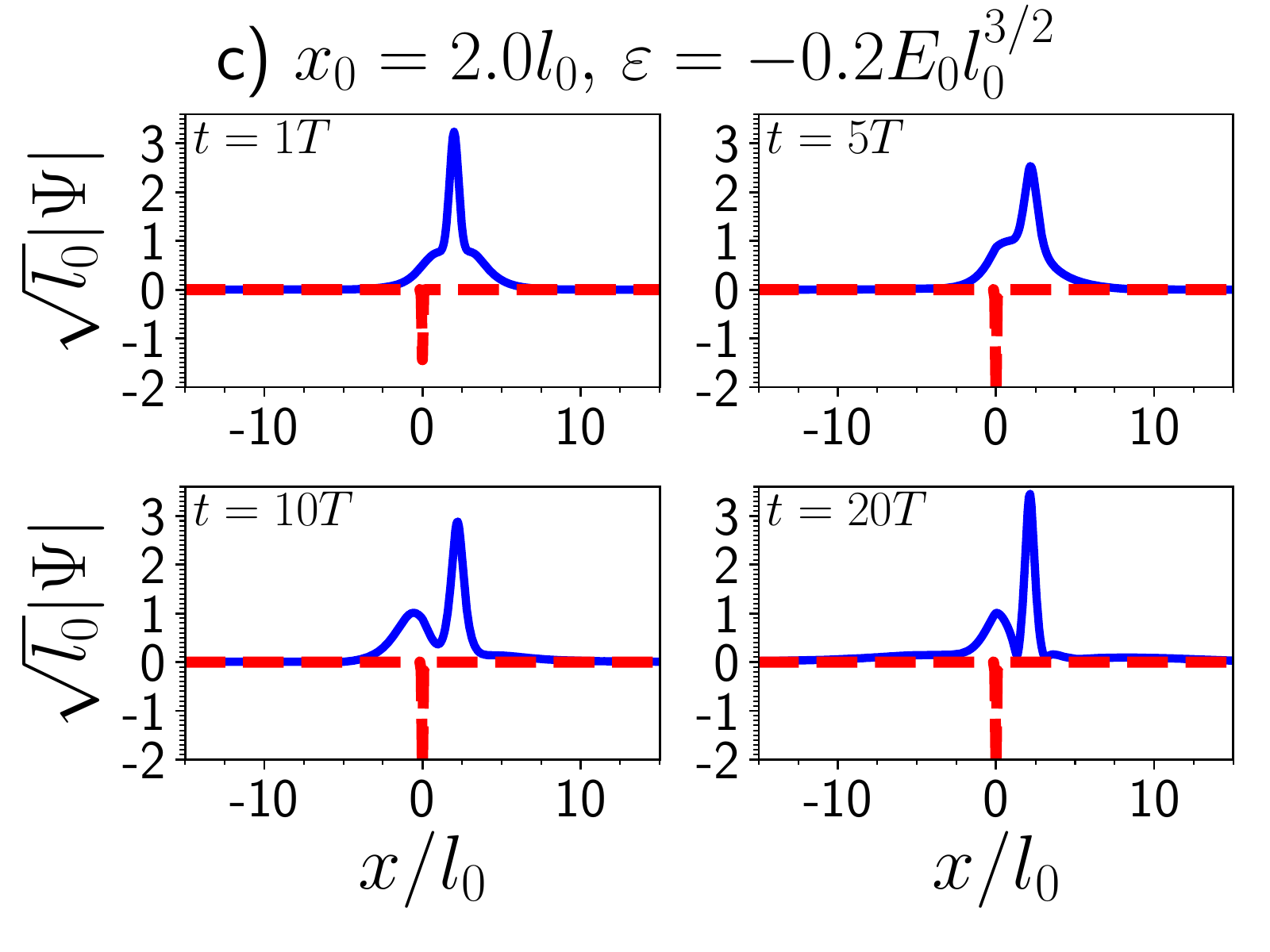}
}\subfloat{\includegraphics[width=0.250\textwidth]{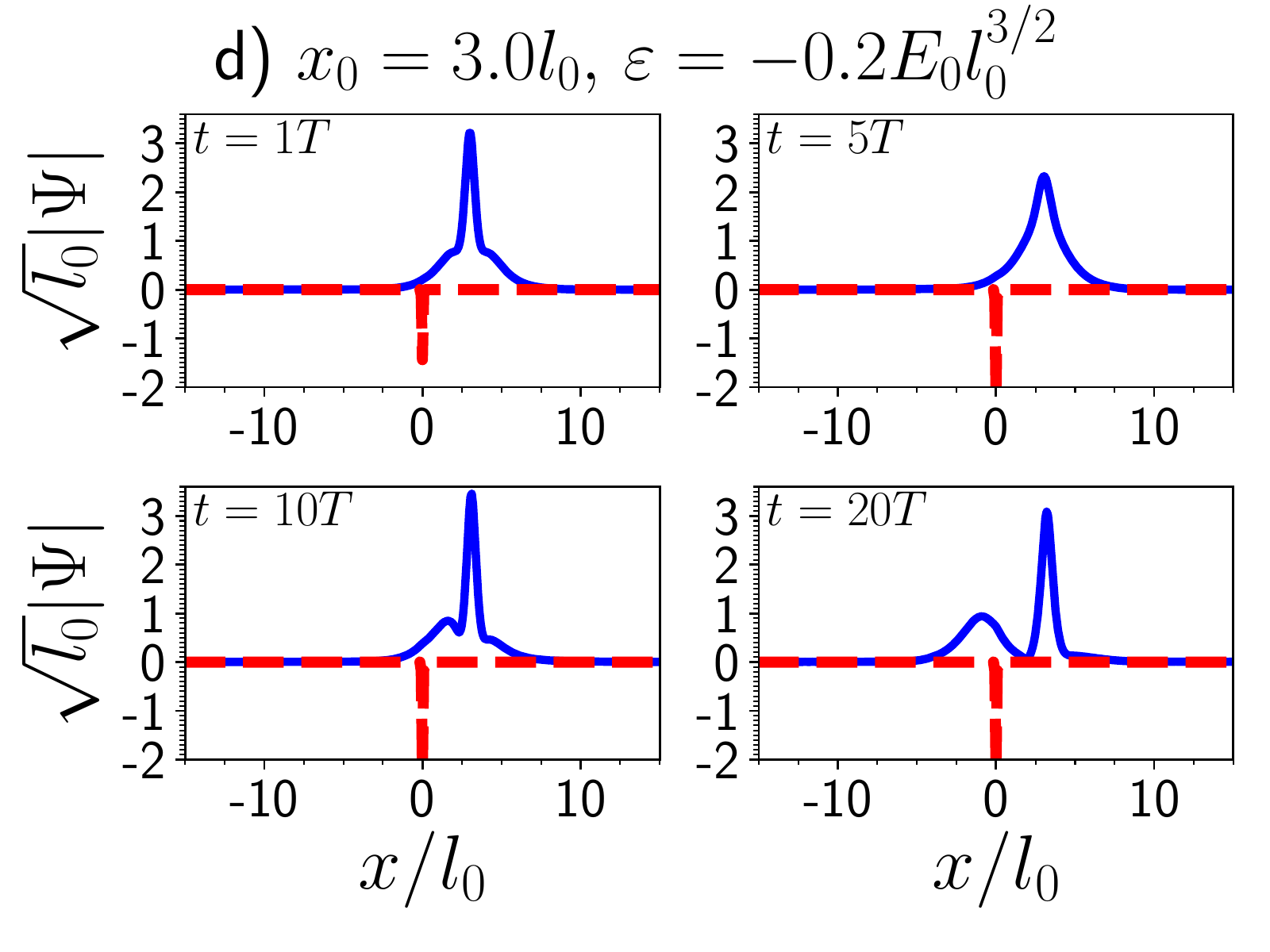}
}
\caption{The same as in Fig. \protect\ref{dir_sim}, but for the splitting of
the breather under the action of the linear potential well, at values of $%
\protect\varepsilon <0$, $x_{0}$, and $t$ indicated in panels.}
\label{att_dir_sim}
\end{figure}

As mentioned above, our goal is to investigate the fission of the breather
into fragments, which will be mapped in the parameter plane of $\left(
\varepsilon ,x_{0}\right) $, for both $\varepsilon >0$ and $\varepsilon <0$,
i.e., in the cases of the delta-like potential barrier and well. We find a
certain threshold value of $x_{0}^{\mathrm{(crit)}}(\varepsilon >0)$ such
that, at $|x|>x_{0}^{\mathrm{(crit)}}$ the breather does not split but
bounces back as a whole from the potential barrier. In the next Section we
demonstrate an analytical estimate for $x_{0}^{\mathrm{(crit)}}$, and then
produce its numerically found values. Note that $x_{0}=0$ is a special
situation that leads to a \emph{spontaneous symmetry breaking} problem, in
the form of spontaneous selection of the directions in which the two major
fragments, which are close to solitons (\ref{sol}), move after completion of
the fission. In numerical calculations, the symmetry-broken outcome may be
determined by a weak random perturbation, if it is added to the input,
simulating the real physical noise, either environmental or quantum.
Numerical noise, induced by truncation error, leads to the symmetry breaking
too, although its characteristics may differ from those of the physical
noise. In this work we demonstrate that even a tiny shift of the input, such
as $|x_{0}|=0.0002$, leads to the apparent symmetry breaking and fission of
the breather, although the fission initiated by very small $|x_{0}|$
develops slowly, see Figs.~\ref{dir_sim}a) and~\ref{dir_sim}b).

\textit{The nonlinear barrier.} It is also possible to consider the
nonlinear potential barrier, based on equation
\begin{equation}
i\frac{\partial \Psi }{\partial t}=-\frac{1}{2}\frac{\partial ^{2}\Psi }{%
\partial x^{2}}-\left[ 1-\frac{\varepsilon _{1}}{\sqrt{\pi }a}f(t)\exp
\left( -\frac{x^{2}}{a^{2}}\right) \right] |\Psi |^{2}\Psi ,  \label{nonlin}
\end{equation}%
where $\varepsilon _{1}$ is the strength of the nonlinear potential. A
permanent localized nonlinear potential was considered in Ref. \cite{HS},
for splitting of an incident fundamental soliton, but not of a breather. The
experimental realization of the nonlinear barrier in an atomic BEC was
proposed in Ref.~\cite{HS}, assuming that a narrow laser beam is applied in
a narrow region, reversing the local interaction between atoms from
attractive to strongly repulsive.
\begin{figure}[tbh]
\centering
\subfloat{\includegraphics[width=0.250\textwidth]{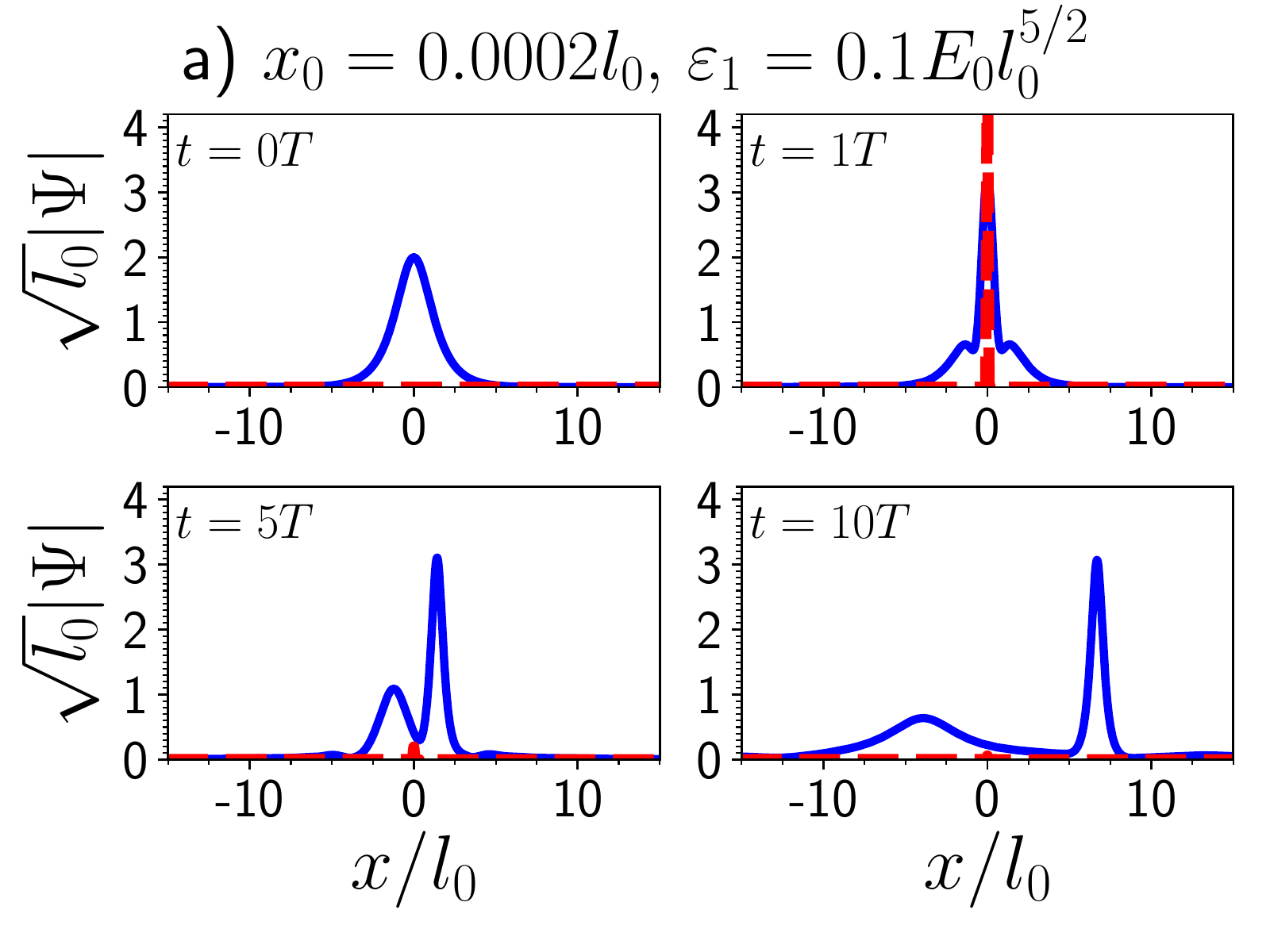}}~%
\subfloat{\includegraphics[width=0.250\textwidth]{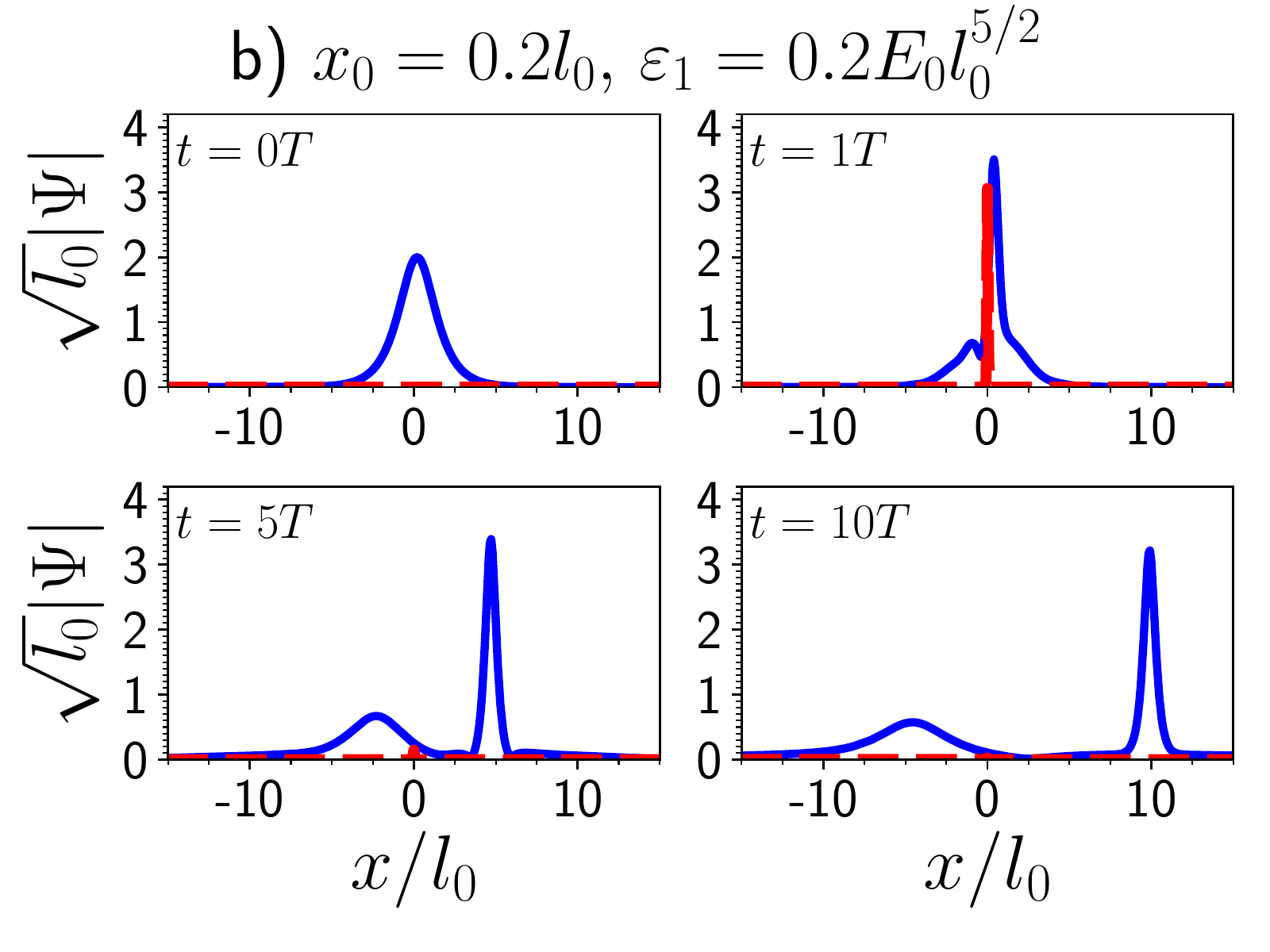}} ~ \\[%
-4ex]
\subfloat{\includegraphics[width=0.250\textwidth]{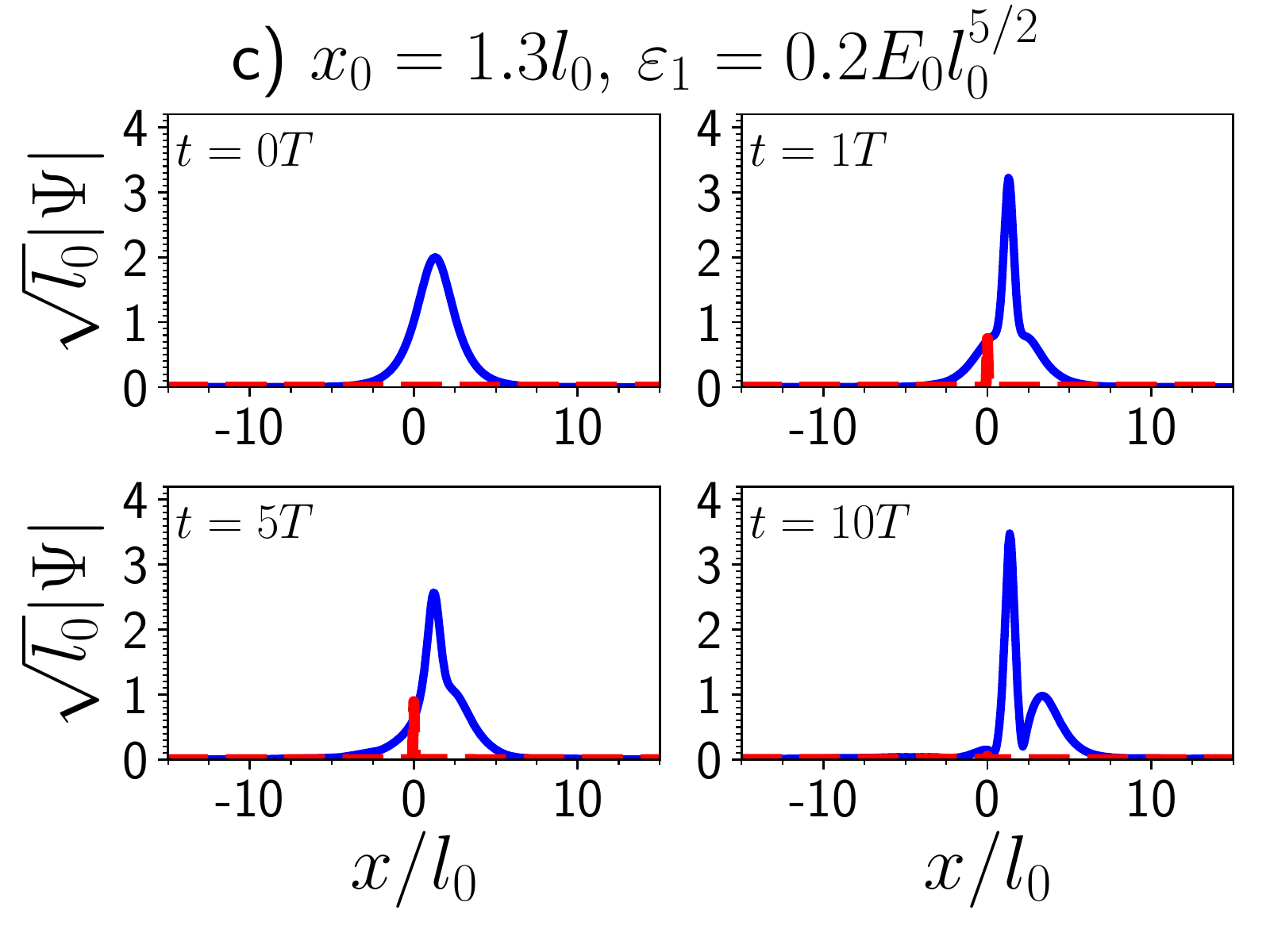}} %
\subfloat{\includegraphics[width=0.250\textwidth]{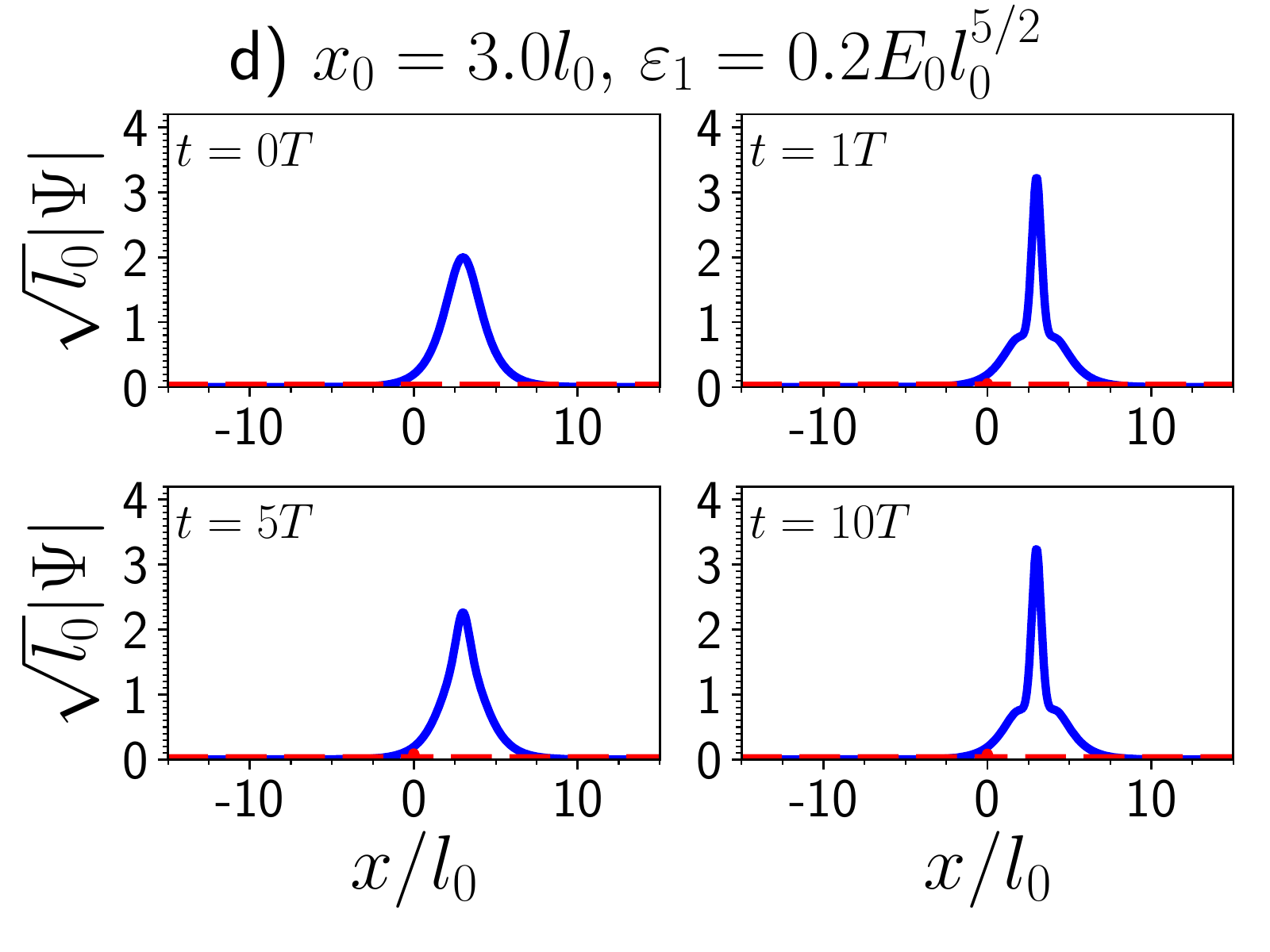}} ~
\caption{The same as in Fig. \protect\ref{dir_sim}, but for the nonlinear
splitter, defined as per Eq. (\protect\ref{nonlin}). In this case, the red
dashed line shows the shape of the splitting barrier, which includes factor $%
\left\vert \protect\psi \left( x=0\right) \right\vert ^{2}$, therefore the
barrier disappears after the separation of the fragments.}
\label{nl_dir_sim}
\end{figure}

\section{\label{sec:methods}Numerical methods and analytical estimates}

The evolution of the breather under the action of the splitting potential
was simulated in the framework of Eq. (\ref{nls}) with $g=1$, using input (%
\ref{t=0}), in which the remaining scaling invariance makes it possible to
fix $A\equiv 1$, while keeping $\varepsilon $ and $x_{0}$ as free
parameters. The simulations were carried out by means of the pseudospectral
method to evaluate spatial derivatives, in a combination with the
Runge-Kutta time-stepping scheme, to secure high accuracy of the results~%
\cite{Y}.

Proceeding to analytical estimates, one can evaluate the threshold value of
the initial displacement, $x_{0}^{\mathrm{(crit)}}$, that separates the
regimes of the bidirectional motion of fragments and unidirectional motion
of the breather as a whole. In the case of large positive $\varepsilon $ and
$\tau =0$ in Eqs. (\ref{nls}) and (\ref{f}), one may assume, in the simplest
approximation, that the input is cut by the tall barrier in two
noncommunicating parts. In particular, the input located at $x<0$ may be
defined as%
\begin{equation}
\psi _{0}(x)=\left\{
\begin{array}{c}
2A~\mathrm{sech}\left[ A\left( x-x_{0}\right) \right] ,~~\mathrm{at}~~x<0,
\\
0,~\mathrm{at}~~x>0.%
\end{array}%
\right. ~  \label{psi0}
\end{equation}%
To estimate whether input \eqref{psi0} may or may not produce a soliton
escaping to the left, it is possible to use the necessary condition for the
generation of at least one NLS soliton in free space by a given input \cite%
{zakharov1984}:%
\begin{equation}
\int_{-\infty }^{+\infty }\left\vert \psi _{0}(x)\right\vert dx>\ln \left( 2+%
\sqrt{3}\right) \approx 1.32.  \label{1.32}
\end{equation}%
Roughly speaking, Eq.~(\ref{1.32}) may be understood as a threshold
condition for the linear scattering problem associated with the NLS
equation, i.e. the Zakharov-Shabat equations~\cite{zakharov1972}, to have a
bound state in the respective potential well. This is a difference from the
linear Schr\"{o}dinger equation, in which, as it is commonly known, an
arbitrarily shallow potential well always maintains at least one bound
state. If input~\eqref{psi0} does not satisfy condition~\eqref{1.32}, it
cannot generate a left-running soliton. Substituting expression \eqref{psi0}
in Eq. \eqref{1.32} predicts a rough estimate for the threshold value $%
x_{0}^{\mathrm{(crit)}}$ which is a boundary for the generation of the
left-traveling soliton:%
\begin{equation}
A~x_{0}^{\mathrm{(crit)}}=\ln \left( \frac{4}{\ln \left( 2+\sqrt{3}\right) }%
\right) \approx 1.11.  \label{thr}
\end{equation}

Additional analytical results may be produced by the conservation of the
Hamiltonian (energy) corresponding to Eq. (\ref{nls}) (with $g=1$) at $%
t>\tau $, i.e., with $f=1$:%
\begin{equation}
E=\frac{1}{2}\int_{-\infty }^{+\infty }\left( \left\vert \frac{\partial \psi
}{\partial x}\right\vert ^{2}-|\psi |^{4}\right) dx+\varepsilon \left\vert
\psi \left( x=0\right) \right\vert ^{2}.  \label{H}
\end{equation}%
The energy of the free-space fundamental solitons (\ref{sol}) contains the
negative ground-state and positive kinetic terms, the effective mass of the
soliton being $2A_{1,2}$ \cite{zakharov1984}:%
\begin{equation}
\left( E_{\mathrm{sol}}\right) _{1,2}=-\frac{1}{3}%
A_{1,2}^{3}+A_{1,2}c_{1,2}^{2}~.  \label{Esol}
\end{equation}%
Note that, in the case of $\varepsilon =0$, the energy of the breather
[the expectation value (\ref{H}) for the solution (\ref{exact})],
\begin{equation}
E_{\mathrm{br}}=-(28/3)A^{3},  \label{Ebr}
\end{equation}%
is exactly equal to the sum of energies (\ref{Esol}) with amplitudes taken
as per Eq. (\ref{A1A2}), and $c_{1,2}=0$. In other words, this simple result
corroborates the known fact that the binding energy of the breather,
considered as a composite state of the constituent solitons, is exactly
zero, in the framework of the integrable NLS equation. For given norm, the
ground state of any physical setting modeled by the NLS equation is the
fundamental soliton. In this connection, it is relevant to stress that the
breather is not a ground state, but rather an excited one. Indeed, using
Eqs. (\ref{Esol}) and (\ref{Ebr}), it is easy to check that the energies of
both states with equal norms are negative, with ratio $E_{\mathrm{sol}}/E_{%
\mathrm{br}}=16/7$.

Assuming that the breather has split in two fundamental solitons with these
amplitudes and velocities $c_{1,2}\neq 0$, the energy conservation, taking
into regard the term $\sim \varepsilon >0$ in Eq. (\ref{H}) and neglecting a
radiative component which may also appear in the output, yields the
following constraint for the velocities:%
\begin{equation}
3c_{1}^{2}+c_{2}^{2}=4\varepsilon A\mathrm{sech}^{2}\left( Ax_{0}\right) .
\label{balance}
\end{equation}

Furthermore, in the case of $x_{0}=0$, the symmetry of the input conserves
the total momentum (otherwise, the splitting potential breaks the
conservation), which implies that the velocities of the fragments are
related by constraint $3c_{1}+c_{2}=0$. In this case, Eq. (\ref{balance})
predicts the individual velocities:%
\begin{equation}
c_{1}=-c_{2}/3=\pm \sqrt{\varepsilon A/3},  \label{c1c2}
\end{equation}%
where $\pm $ implies that the spontaneous symmetry breaking may randomly
send the soliton with velocity $c_{1}$ in either direction.

If a sufficiently strong perturbation transforms the initial breather (\ref%
{t=0}) into a pair of fundamental solitons with the conservation of the
total norm, i.e., $A_{1}+A_{2}=4A$, but the amplitude ratio $A_{1}:A_{2}$
different from $3:1$ [see, e.g., Fig. \ref{nl_dir_sim}(a)], the comparison
of energies (\ref{Esol}) and (\ref{Ebr}) demonstrates that the energy of the
soliton pair with zero velocities is larger than the initial energy in the
case of $1<A_{1}:A_{2}<3$ (hence, this conversion scenario is suppressed),
and is smaller in the cases of%
\begin{equation}
A_{1}:A_{2}<1~~\mathrm{or~~}A_{1}:A_{2}>3,~  \label{ratio}
\end{equation}%
facilitating these scenarios.

In the model with a nonlinear splitting potential, which corresponds to Eq. (%
\ref{nonlin}), similar analytical predictions are produced by replacement $%
\varepsilon \rightarrow 2A^{2}\varepsilon _{1}$, i.e.,%
\begin{gather}
3c_{1}^{2}+c_{2}^{2}=8\varepsilon _{1}A^{3}~\mathrm{sech}^{2}\left(
Ax_{0}\right) ,  \label{balance_2} \\
c_{1}=-c_{2}/3=\pm \sqrt{2\varepsilon _{1}A^{3}/3}~.  \label{c1c2_2}
\end{gather}

\section{\label{sec:results}Results}

The evolution of the breather interacting with the local potential is
reported in Figs. \ref{dir_sim}, \ref{att_dir_sim}, and \ref{nl_dir_sim} by
showing snapshots of numerically generated solutions to Eqs.~\eqref{nls} and~%
\eqref{nonlin}, taken at different times. 
Further, the results are summarized below in Fig. \ref{lin_heatmap} and, 
additionally, in Appendix (Figs. \ref%
{nlb_heatmap}-\ref{nl_vel2}) by means of color-coded plots, which display
amplitudes and velocities of the fragments, in the plane of $\left(
\varepsilon ,x_{0}\right) $ or $\left( \varepsilon _{1},x_{0}\right) $, for
the linear and nonlinear splitting potentials, respectively.

\subsection{\label{sec:lin_results}The linear potential barrier or well}

In this subsection we present results obtained by solving Eq.~\ref{nls} with
the linear potential, defined according to Eqs.~\eqref{U} and~\eqref{f}, and
the input taken as per Eq.~\eqref{exact}. The simulations reveal two main
patterns of behavior: (i) the breather splits in two pulses close to
fundamental solitons with the amplitude ratio close to $3:1$, cf. Eq. (\ref%
{sol}), moving in opposite directions; or (ii) the breather is pushed by the
barrier to one side, again splitting in two solitons, subject to the same
approximate amplitude ratio, $3:1$, but moving in one direction. Figure~\ref%
{dir_sim} displays snapshots of the simulations for different values of $%
x_{0}$ and $\varepsilon $, which illustrate these dynamical scenarios.

\begin{figure}[th]
\centering
\includegraphics[width=0.5\textwidth]{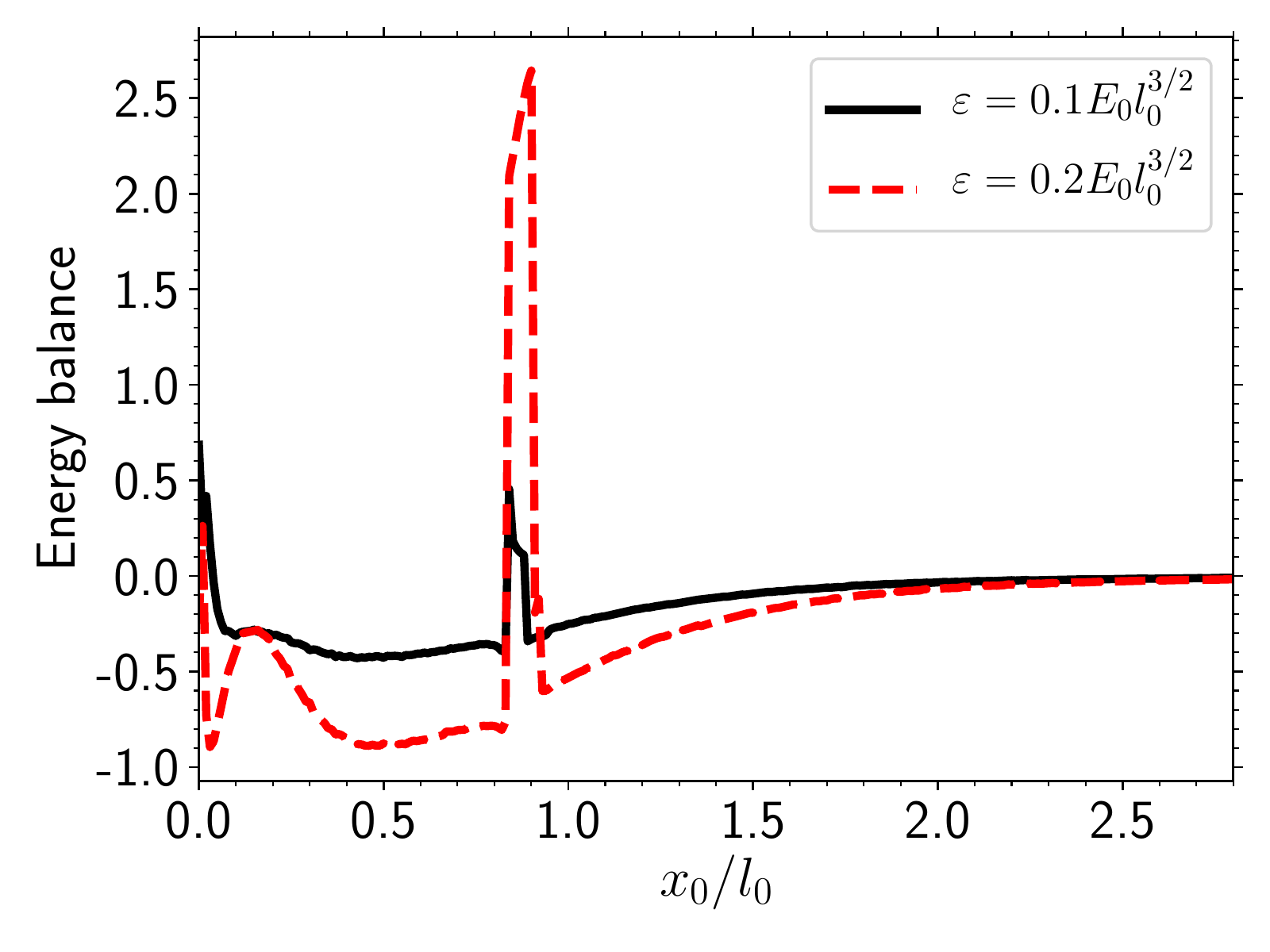}
\caption{The difference of the left- and right-hand sides of Eq.~
\eqref{balance}, as obtained from numerical data, illustrating the accuracy
of the analytically predicted energy balance, for two values of the
linear-barrier's height, $\protect\varepsilon =0.1$ and $0.2$.}
\label{fig_cond17}
\end{figure}

In Fig.~\ref{fig_cond17} we show values of $3c_{1}^{2}+c_{2}^{2}-4%
\varepsilon A^{2}\mathrm{sech}^{2}(Ax_{0})$ for $\varepsilon =0.1$ and $0.2$%
, to demonstrate accuracy of the analytical energy-balance constraint,
predicted by Eq.~\eqref{balance}, in comparison with the numerical results.
Velocities of the fragments are evaluated numerically as
\begin{equation}
c_{j}=\frac{x_{j}(t_{\max })-x_{j}(t_{\max }/2)}{t_{\max }/2},
\label{vel_def}
\end{equation}%
where $x_{j},~j=1,2$, are positions of the $j$-th peak, taken at times $%
t_{\max }=15T$ and $t_{\max }/2=7.5T$, to ensure that the peaks have
separated (in most cases). It is seen that the predicted energy balance
condition holds well for the two fragments that move in the same direction,
but is completely broken in a vicinity of the critical point $x_{0}^{(%
\mathrm{crit})}$, as expected. For the two fragments moving in opposite
directions, the energy balance is not satisfied very accurately, but the
discrepancy is relatively small. Close to to the symmetry-breaking point, $%
x_{0}=0$, Fig.~\eqref{fig_cond17} shows that the analytically predicted
condition is not satisfied either, due to the slow separation of the
fragments, which means that the ballistic regime for the fragments was not
yet achieved.
\begin{figure}[th]
\centering
\includegraphics[scale=0.5]{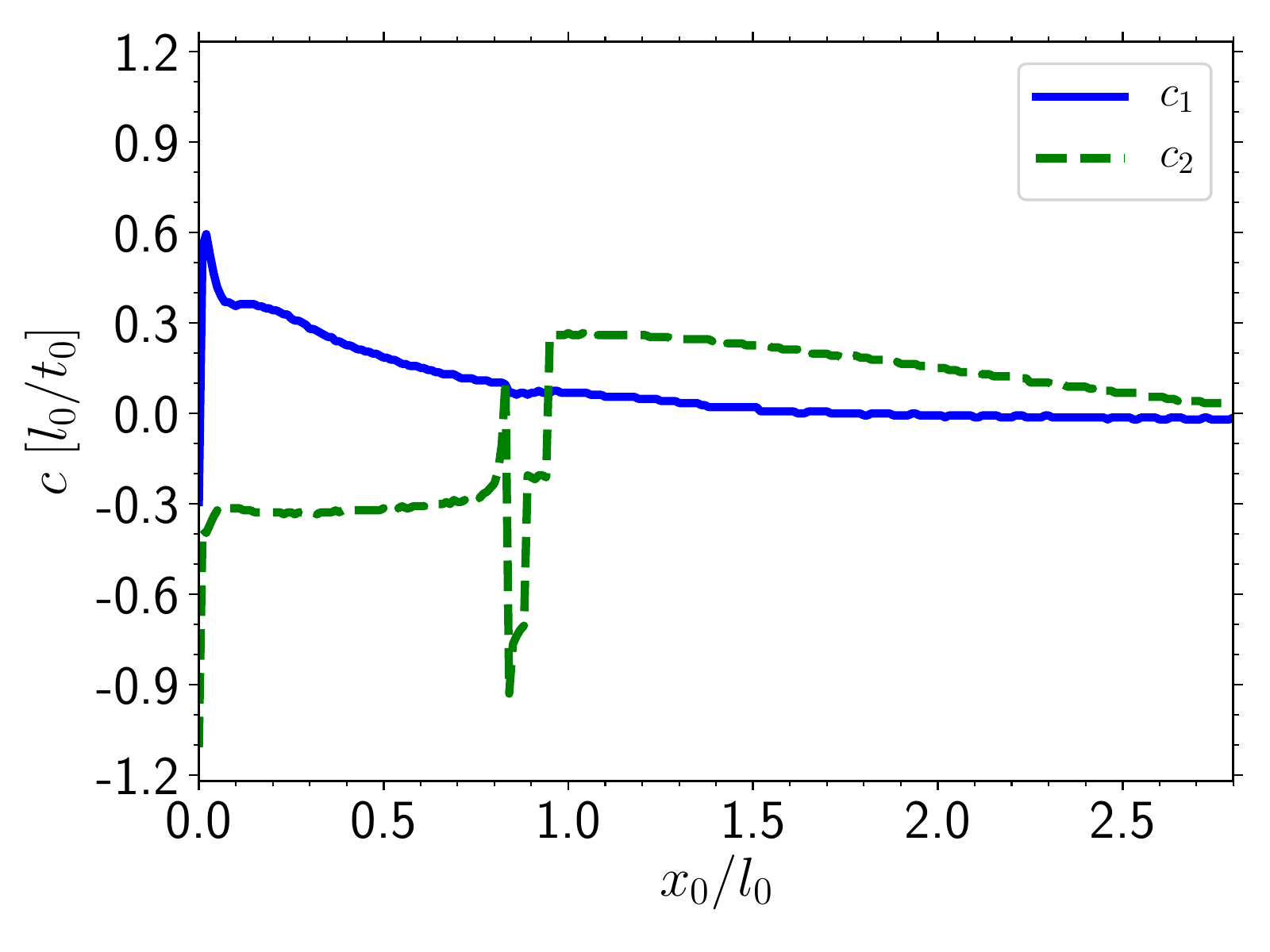} %
\includegraphics[scale=0.5]{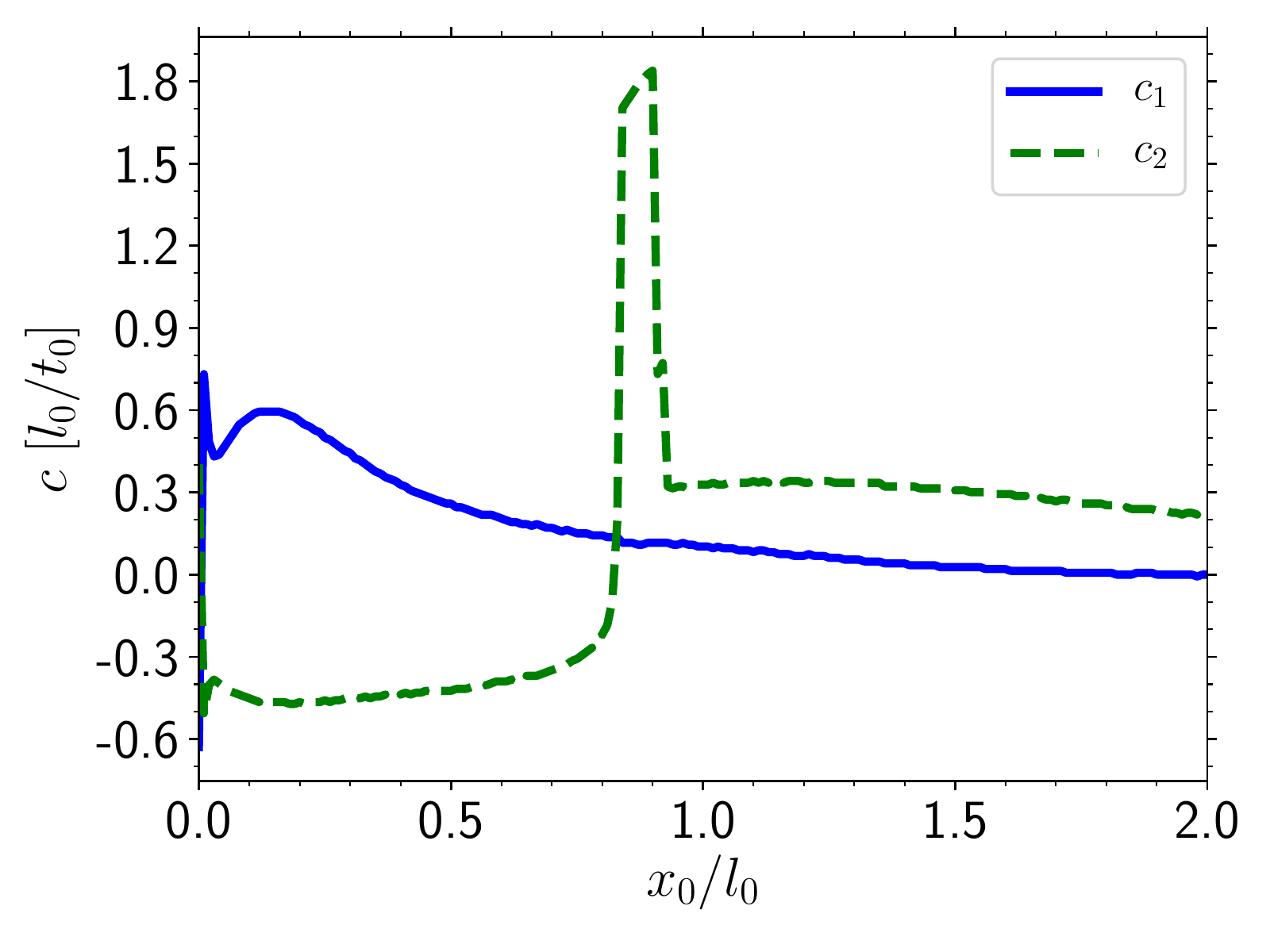}
\caption{Velocities of the fragments, numerically identified according to
Eq.~ \eqref{vel_def}, as functions of $x_{0}$ for fixed values of the
linear-barrier's height, $\protect\varepsilon =0.1$ and $\protect\varepsilon %
=0.2$. The velocity of the heavier and lighter fragments, $c_{1}$ and $c_{2}$%
, are shown by the blue solid and green dashed curves, respectively}
\label{velocities_lin}
\end{figure}
In Figs.~\ref{velocities_lin} we show the velocities of the fragments
according to the definition in Eq.~\eqref{vel_def} as functions of $x_{0}$
for fixed values of the barrier height, $\varepsilon =0.1$ and $\varepsilon
=0.2$.

As mentioned above, point $x_0 = 0$ is a special one. Indeed, due to the
symmetry of input Eq. (\ref{t=0}), the breather cannot deterministically
split into the solitons with unequal amplitudes. However, the symmetry may
be spontaneously broken by an external perturbation. In Figs~\ref{dir_sim}a)
and~\ref{dir_sim}b) we show that even a tiny change of the initial
conditions leads to the symmetry breaking and, eventually, fission of the
breather.

Typical examples of the interaction of the breather with the potential well (%
$\varepsilon <0$), displayed in Fig. \ref{att_dir_sim}, demonstrate that the
breather does not split in the limit of small initial displacements, $x_{0}\rightarrow 0$. In this case
[panel a) in the figure], the breather sheds off some radiation, and
transforms into a fundamental soliton pinned to the potential well. When $%
x_{0}$ increases, the breather splits, but the fragments, being pulled by
the attractive potential, separate much slower than in the case of the
repulsive barrier, cf. Fig. \ref{dir_sim}. In particular, the heavier
soliton may stay trapped by the potential well (performing small
oscillations in this state), while the lighter one moves very slowly, as
seen in panel b). This outcome of the interaction can be easily explained by
the energy balance. Indeed, for small $|\varepsilon |$ comparison of the
initial interaction energy, given by the last term in Eq. (\ref{H}), with $%
\left\vert \psi \left( x=0\right) \right\vert ^{2}=4A^{2}\mathrm{sech}%
^{2}\left( Ax_{0}\right) $, and a similar expression for the fundamental
solitons with amplitudes Eq. (\ref{A1A2}), i.e., $\left\vert \psi \left(
x=0\right) \right\vert ^{2}=A_{1,2}^{2}$, demonstrates that the trapped
state is energetically possible for the soliton with the larger amplitude, $%
A_{1}=3A$, in which case the energy-balance equation yields%
\begin{equation}
c_{2}^{2}=|\varepsilon |A\left[ 9-4\mathrm{sech}^{2}\left( Ax_{0}\right) %
\right] ,  \label{c2}
\end{equation}%
cf. Eqs. (\ref{balance}) and (\ref{c1c2}).

\begin{figure}[th]
\centering
\includegraphics[width=0.50\textwidth]{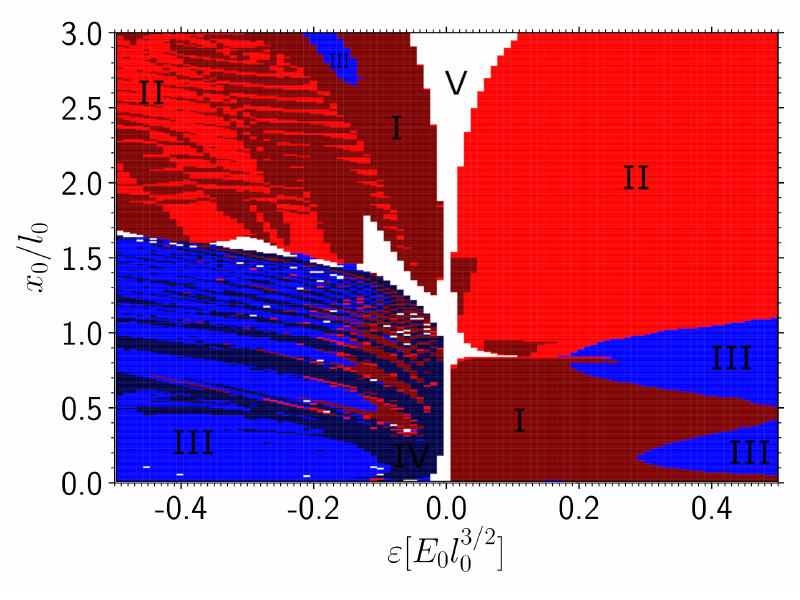}
\caption{\textit{The linear splitter}. The map of the regimes the splitting
of the original breather in simulations of Eq. (\protect\ref{nls}) at $t=15T$%
, in the plane of parameters $\protect\varepsilon $ and $x_{0}$. Different
regimes are color-coded and denoted by Roman numerals: \textbf{I}) the
breather splits in two soliton-like fragments moving in opposite
directions(brown); \textbf{II}) the breather splits in two soliton-like
fragments moving in the same direction (red); \textbf{III}) the breather
does not split clearly enough in the course of the simulation time, $t=15T$
(blue); \textbf{IV}) the breather splits in a heavies soliton-like fragment,
bound by the attractive potential, and radiation (dark blue); \textbf{V})
the breather does not split (white).}
\label{lin_heatmap}
\end{figure}

There are several possible regimes of the breather's evolution following the
interaction with the external potential: i) the initial breather splits in
two soliton-like pulses that move bidirectionally; ii) the breather splits
in two pulses that move unidirectionally; iii) the breather does not split
clearly enough to distinguish the soliton-like fragments after the evolution
time of the simulation; iv) the breather is transformed into a soliton bound
by the trapping well; v) the breather does not split. In Fig.~\ref%
{lin_heatmap} we map these regimes in the parameters plane of $\left(
\varepsilon ,x_{0}\right) $. They are color-coded and denoted with Roman
numerals. To plot the map in Fig.~\ref{lin_heatmap} we ran the simulations
up to $t=15T$, and used the breather's width, $l_{\mathrm{br}}=1/A$ , as a
measure of the separation of the fragments. Thus, we categorize the breather
as a split state when the distance between center-of-mass of the fragments
is equal to or exceeds $l_{\mathrm{br}}$.

In Appendix we present the results for amplitudes and velocities of the two
largest peaks in solution density at $t=15T$, in the form of color-coded
maps in the plane of $\left( \varepsilon ,x_{0}\right) $.

\subsection{The nonlinear potential barrier}

\begin{figure}[th]
\centering
\includegraphics[width=0.5\textwidth]{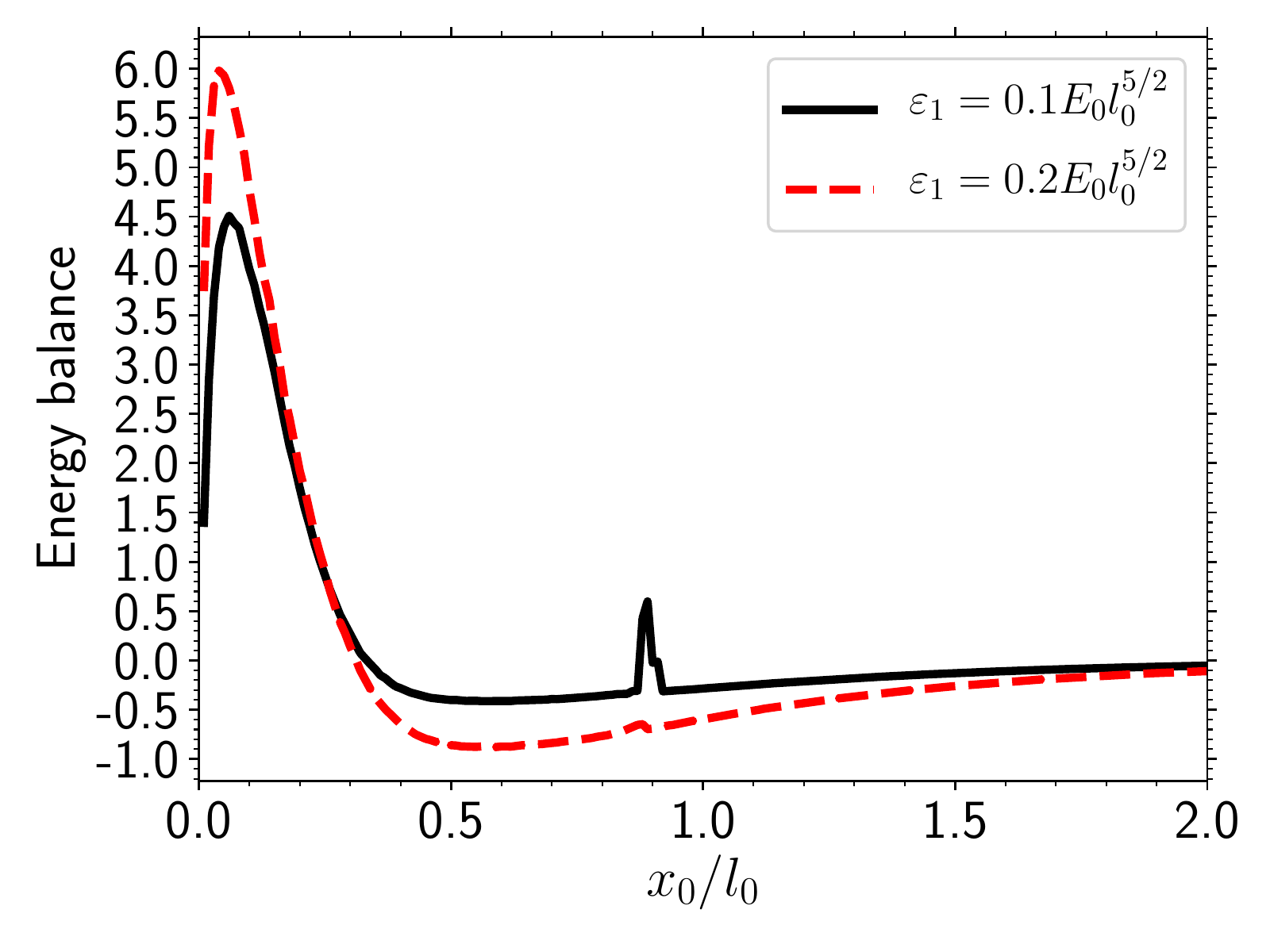}
\caption{The same as in Fig. \protect\ref{fig_cond17}, but for the energy
balance predicted by Eq.~\eqref{balance_2} for two values of the nonlinear
barrier height, $\protect\varepsilon =0.1$ and $0.2$.}
\label{fig_cond20}
\end{figure}

\begin{figure}[th]
\centering
\includegraphics[scale=0.5]{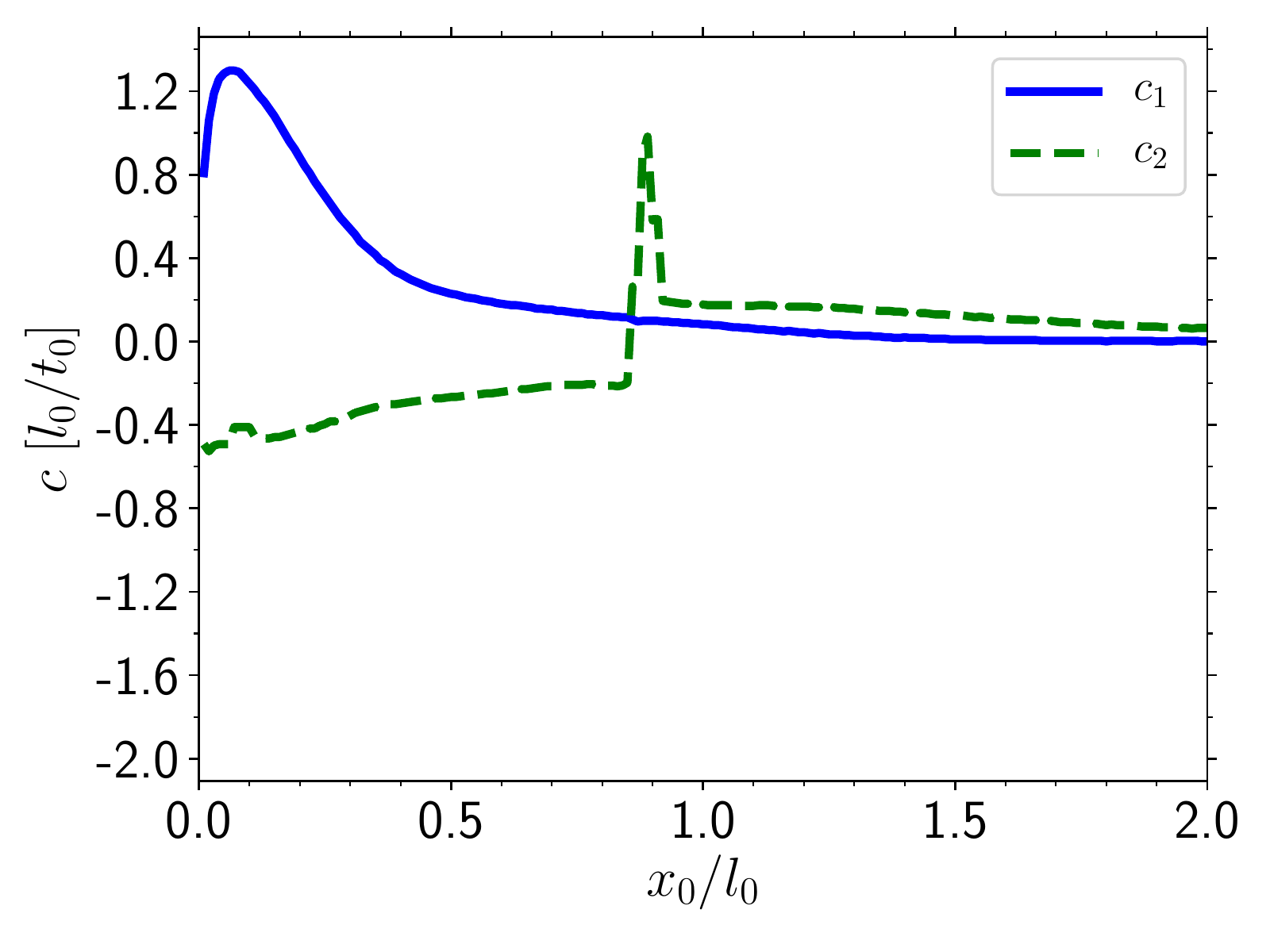} %
\includegraphics[scale=0.5]{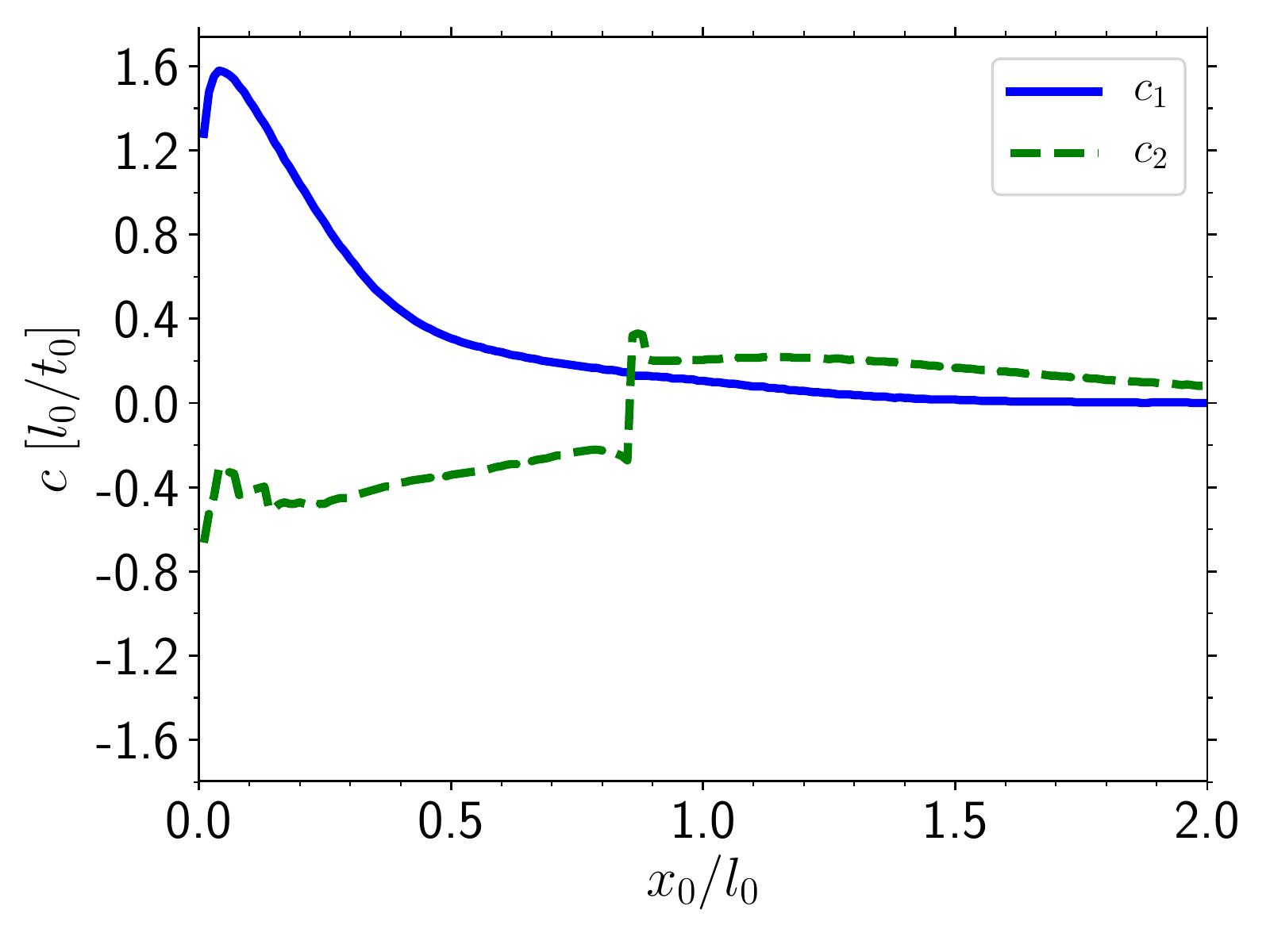}
\caption{Velocities of the fragments, identified as per Eq. \eqref{vel_def},
as functions of $x_{0}$ for fixed values of the nonlinear-barrier's height, $%
\protect\varepsilon_1 =0.1$ and $\protect\varepsilon =0.2$. The velocities
of the heavier and lighter fragment, $c_{1}$ and $c_{2}$, are shown by the
blue solid and green dashed curves, respectively.}
\label{velocities_nlb}
\end{figure}

In the case of the nonlinear potential defined by Eq.~\eqref{nonlin}, the
difference from its linear-potential counterpart is that, as the strength of
the nonlinear barrier depends on the local density at $x=0$, the action of
the effective potential subsides in the course of splitting. Typical
examples of the interaction of the breather with the nonlinear splitter are
presented in Fig.~\ref{nl_dir_sim}. As in the case of the linear barrier,
even a tiny asymmetry in the initial conditions, with $|x_{0}|=0.0002$,
breaks the symmetry of the problem and leads to fission of the breather. The
resulting amplitude ratio is $\simeq 10:1$, which is much larger than the
\textquotedblleft natural" one, $3:1$.
In Fig.~\ref{fig_cond20} we show the accuracy of the energy balance for the
nonlinear barrier, as predicted by Eq.~\eqref{balance_2}. We see that,
similarly to the linear barrier, the predicted condition is very accurate
for the fragments moving in the same direction, and is broken in a vicinity
of the critical point $x_{0}^{(\mathrm{crit})}$, as expected. However, the
breaking of the predicted energy-balance condition is much more pronounced
in the case of the nonlinear potential, as seen in the $x_{0}<x_{0}^{(%
\mathrm{crit})}$ segment of Fig.~\ref{fig_cond20}.
In Figs.~\ref{velocities_nlb} we plot the velocities of the two largest
density peaks following the interaction of the input with the nonlinear
barrier, that are identified as per Eq.~\eqref{vel_def}.

\begin{figure}[th]
\centering
\includegraphics[width=0.50\textwidth]{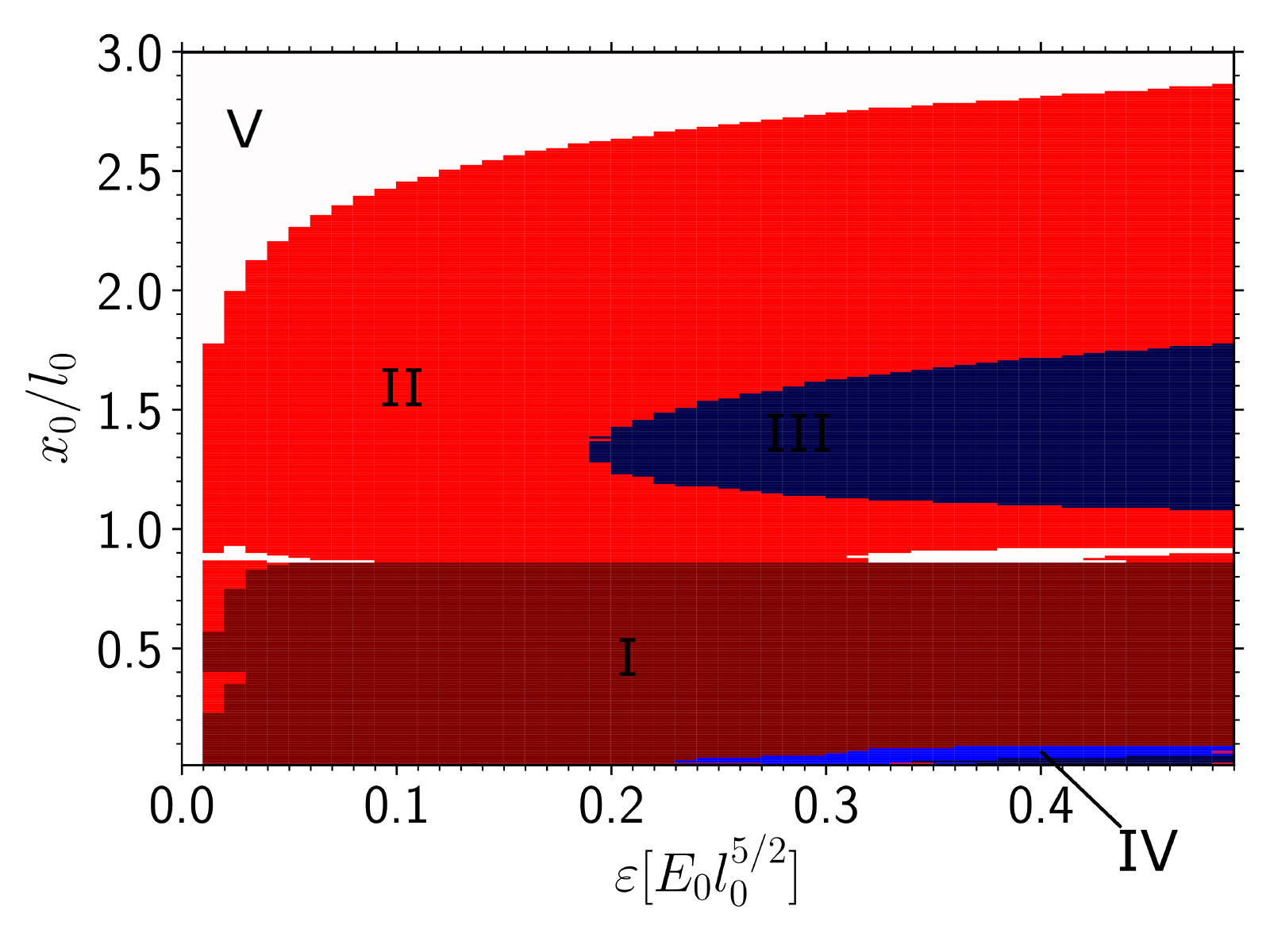}
\caption{\textit{The nonlinear splitter.} The map of splitting regimes for
the original breather in simulations of Eq. (\protect\ref{nonlin}) at $t=15T$%
, in the plane of parameters $\protect\varepsilon _{1}$ and $x_{0}$.
Different regimes are color-coded and denoted by Roman numerals: \textbf{I})
the breather splits in two soliton-like fragments moveing in opposite
directions(brown); \textbf{II}) the breather is not separated clearly enough
at $t=15T$ (red); \textbf{III}) the breather splits in two soliton-like
fragments moving in the same direction (dark blue); \textbf{IV}) the
breather splits in one soliton and radiation (blue); \textbf{V}) the
breather does not split. }
\label{nlb_heatmap}
\end{figure}

Just as for the linear splitting potential, there are several possible
regimes of the breather's evolution after the interaction with the nonlinear
splitter: i) the initial breather splits into two soliton-like pulses that
move bidirectionally; ii) the breather does not split clearly enough to
distinguish soliton-like fragments after the simulation time; iii) the
breather splits in two soliton-like unidirectionally moving pulses; iv) the
breather splits in a soliton and radiation; v) the breather does not split.
In Fig.~\ref{nlb_heatmap} we map these regimes in the parameters plane of $%
\left( \varepsilon ,x_{0}\right) $. The different regimes are color-coded
and denoted by Roman numerals. For the map in Fig.~\ref{nlb_heatmap} we ran
the simulations up to $t=15T$, and used the width of a breather, $l_{\mathrm{%
br}}=1/A$ , as a measure of the separation of the fragments.

We see that in the case of the nonlinear potential the general character of
the breather's splitting is similar to the case of the linear repulsive
barrier. However, the region where both fragments travel in the same
direction is much smaller, in comparison to the latter case.

The results for amplitudes and velocities of the two largest density peaks
of the solution at $t=15T$, in the form of color-coded maps in the plane of $%
\left( \varepsilon ,x_{0}\right) $, for the setting with the nonlinear
barrier are displayed in Appendix too.

\section{\label{sec:conclusions}Conclusion}

The subject of the work is the dynamics of NLS breathers interacting with
localized linear and nonlinear splitting potentials. The setting applies to
matter-wave states in a BEC, as well as to optical beams and pulses, in the
spatial and temporal domains, respectively. The potentials are approximated
by the delta-function, and by its regularized version, as concerns the
numerical scheme. The linear splitter may be represented by a narrow barrier
or potential well (the nonlinear splitter is considered only with the
repulsive sign). The potentials breaks the integrability of the NLS
equation, leading to various dynamical scenarios. Depending on the initial
position of the breather with respect to the localized splitting potential,
it may split into a pair of fundamental solitons, moving in opposite
directions, or bounce as a whole. In the former case, the amplitude ratio of
the fragments may be both close to the expected value $3:1$, predicted by
the exact solution for the NLS breather, and strongly different from it
(especially in the case of the nonlinear splitter). When the center of the initial breather is very close to the position of the
barrier's center, the linear potential well leads to another outcome of the
interaction, with the larger-amplitude soliton staying trapped by the
attractive well, while the smaller-amplitude one slowly escapes.
Post-splitting velocities of the solitons may be predicted by means of the
analytical energy balance.

There are several avenues to explore as a follow-up to this work. First and
foremost, one may consider splitting potentials with shapes that are closer
to experimentally realistic ones. For getting closer to the experimentally
relevant setting in BEC, it may also be necessary to consider the full
three-dimensional model, rather than its one-dimensional reduction, cf. Ref.
\cite{we}. Another relevant extension may be inclusion of quantum noise,
again in terms of a BEC. In particular, quantum fluctuations may govern the
spontaneous symmetry breaking in the case of $x_{0}=0$.

\section{Acknowledgement}

This work is jointly supported by the National Science Foundation through
grants No. PHY-1402249 and No. PHY-1607221 and the Binational (US-Israel)
Science Foundation through grant No. 2015616,and by Israel Science
Foundation (project No. 1287/17). The work at Rice was supported by the Army
Research Office Multidisciplinary University Research Initiative (Grant No.
W911NF-14-1-0003), the Office of Naval Research, the NSF (Grant No.
PHY-1707992), and the Welch Foundation (Grant No. C-1133).

\onecolumngrid
\appendix*

\section{Amplitude and velocity heatmaps}

\label{appendixA} In this Appendix we present additional computational
results, namely two-dimensional maps of the amplitudes and velocities of the
heavier and lighter fragments in the settings with both linear and nonlinear
localized potentials.

\subsection{The linear splitter}

\begin{figure}[th]
\centering
\includegraphics[width=0.4\textwidth]{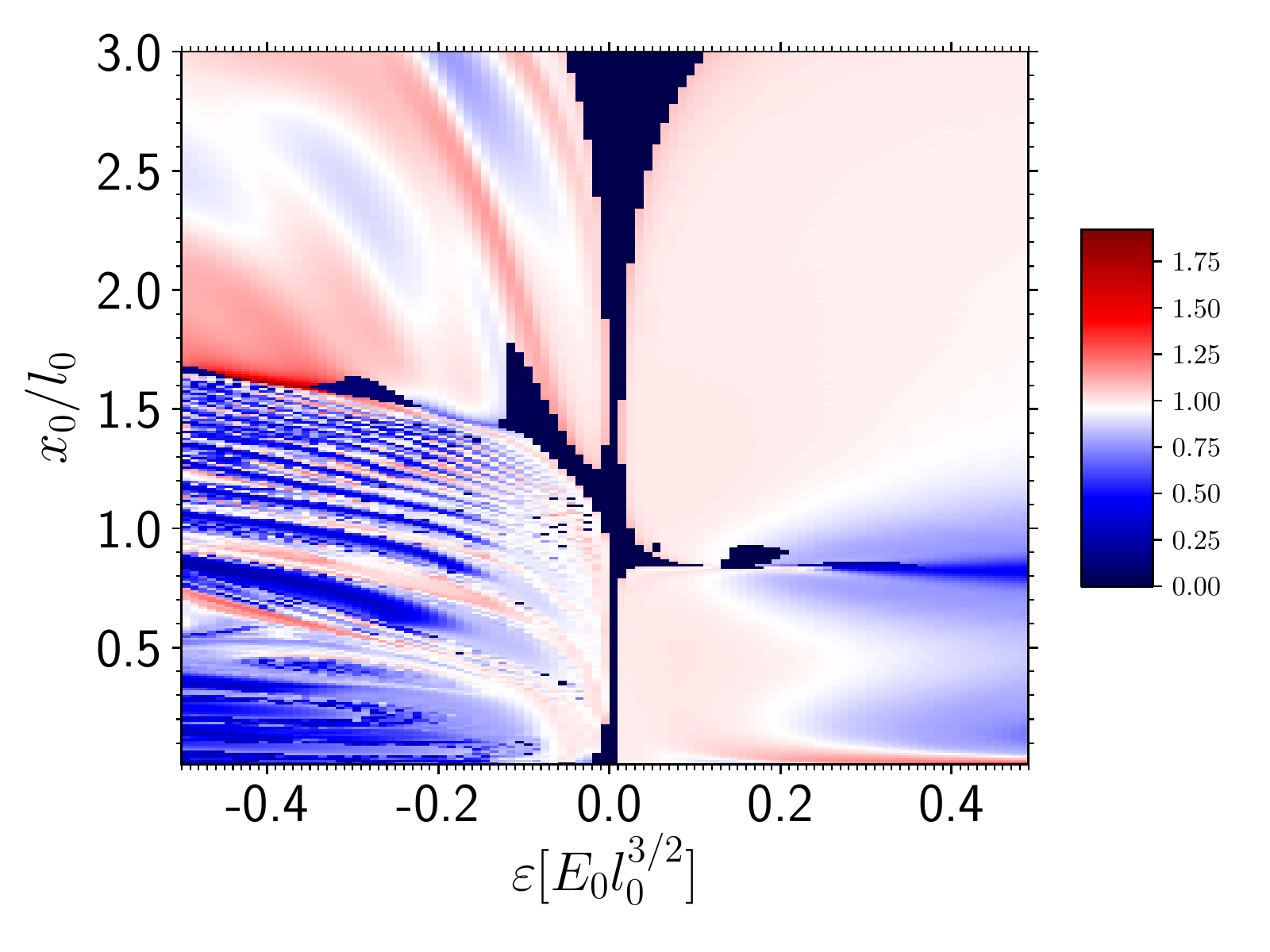}
\caption{\textit{The linear splitter.} The same as in Fig. \protect\ref%
{lin_ampl1}, but for the amplitude of the lighter soliton produced by the
splitting of the original breather.}
\label{lin_ampl2}
\end{figure}

Results for the heavier and lighter fragments are collected in Figs.~\ref%
{lin_ampl1} and \ref{lin_ampl2}, in the form of color-coded maps in the
plane of $\left( \varepsilon ,x_{0}\right) $, which also include the data
produces for $\varepsilon <0$, i.e., for the splitting produced by the
interaction of the breather with the narrow potential well. These figures,
along with Fig. \ref{dir_sim}, clearly demonstrate that, in the majority of
cases in the case of splitting barrier, $\varepsilon >0$, the amplitude
ratio for the fragments remains close to the expected value $3:1$.

\begin{figure}[th]
\centering
\includegraphics[width=0.4\textwidth]{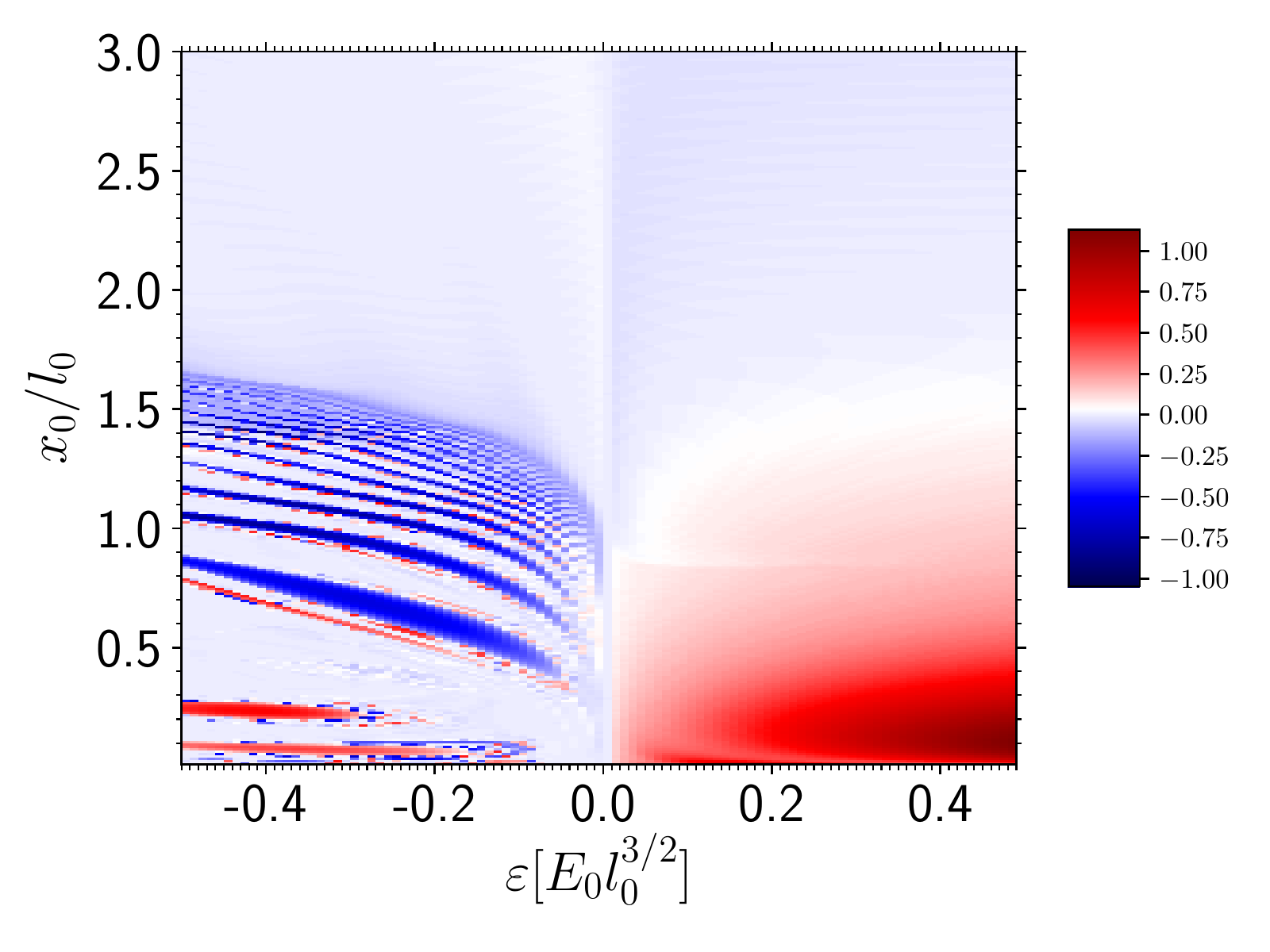}
\caption{\textit{The linear splitter.} The map of values of the velocity of
the heavier soliton, produced by the splitting of the original breather in
simulations of Eq. (\protect\ref{nls}) at $t=15T$, in the plane of
parameters $\protect\varepsilon $ and $x_{0}$. Black dashed lines,
separating different regions in the diagram, serve as a guide to the eye.
The light blue hue for large values of $x_{0}$ corresponds to zero velocity.}
\label{lin_vel1}
\end{figure}

\begin{figure}[th]
\centering
\includegraphics[width=0.40\textwidth]{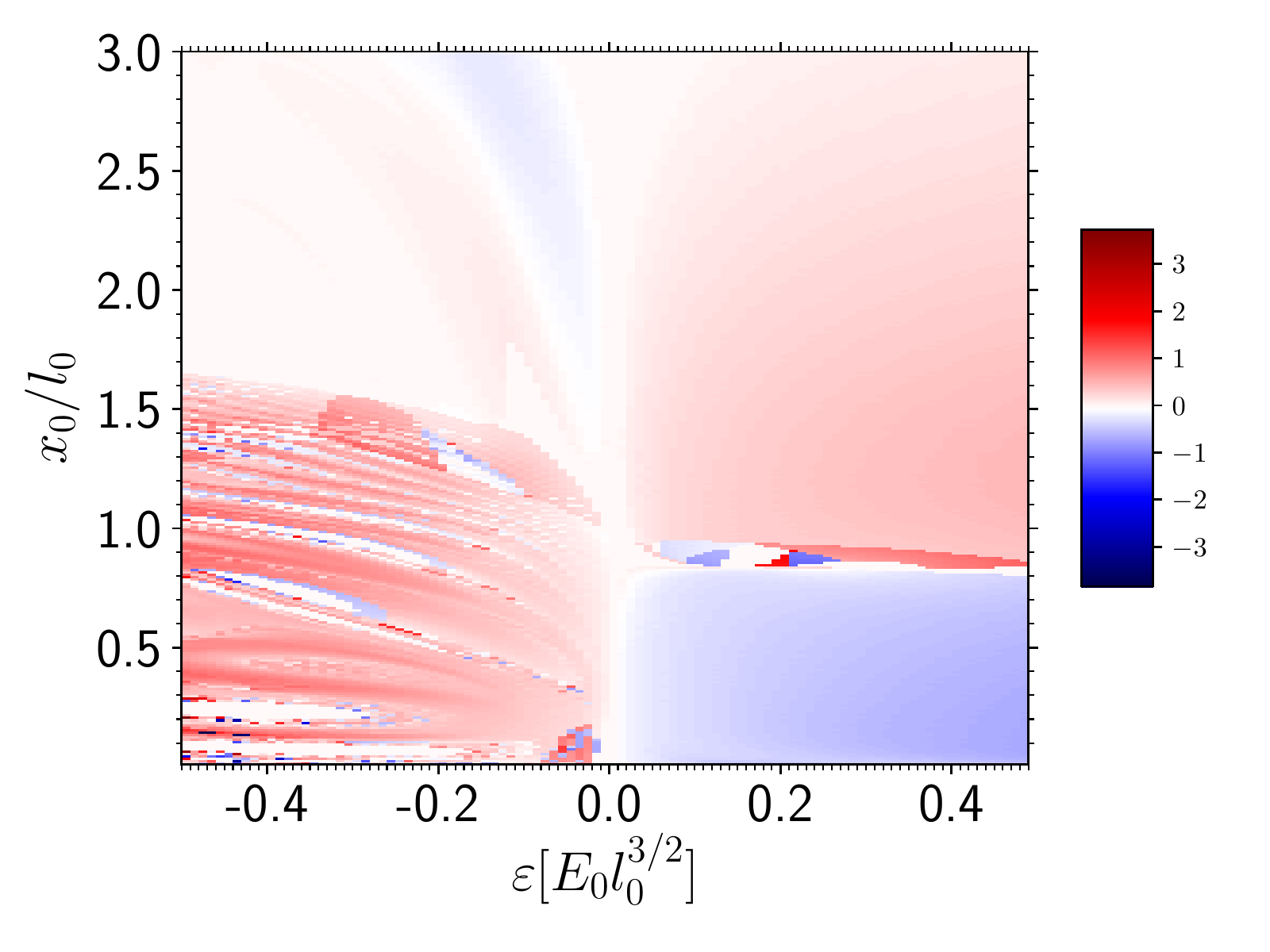}
\caption{\textit{The linear splitter.} The same as in Fig. \protect\ref%
{lin_vel1}, but for the velocity of the lighter soliton (the one with a
smaller amplitude).}
\label{lin_vel2}
\end{figure}

Results for velocities of the fragments are collected in Figs.~\ref{lin_vel1}
and \ref{lin_vel2}. In particular, the former figure demonstrates that the
sign of the velocity of the heavier soliton is always identical to the sign
of $x_{0}$, whereas the it is seen in Fig. \ref{lin_vel2} that the lighter
soliton moves in the opposite direction at
\begin{equation}
\left\vert x_{0}\right\vert <x_{0}^{\mathrm{(crit)}}\approx 0.9,  \label{0.9}
\end{equation}%
and changes the sign of its velocity at $\left\vert x_{0}\right\vert >x_{0}^{%
\mathrm{(crit)}}$, moving in the same direction as its heavier counterpart.
Note that the value of $x_{0}^{\mathrm{(crit)}}$is rather close to the one
predicted above by the crude analytical approximation, see Eq. \eqref{thr}.
Lastly, the numerical results for the velocities approximately corroborate
the analytical prediction produced by Eqs. (\ref{balance}) and (\ref{c1c2}).

\begin{figure}[th]
\centering
\includegraphics[width=0.40\textwidth]{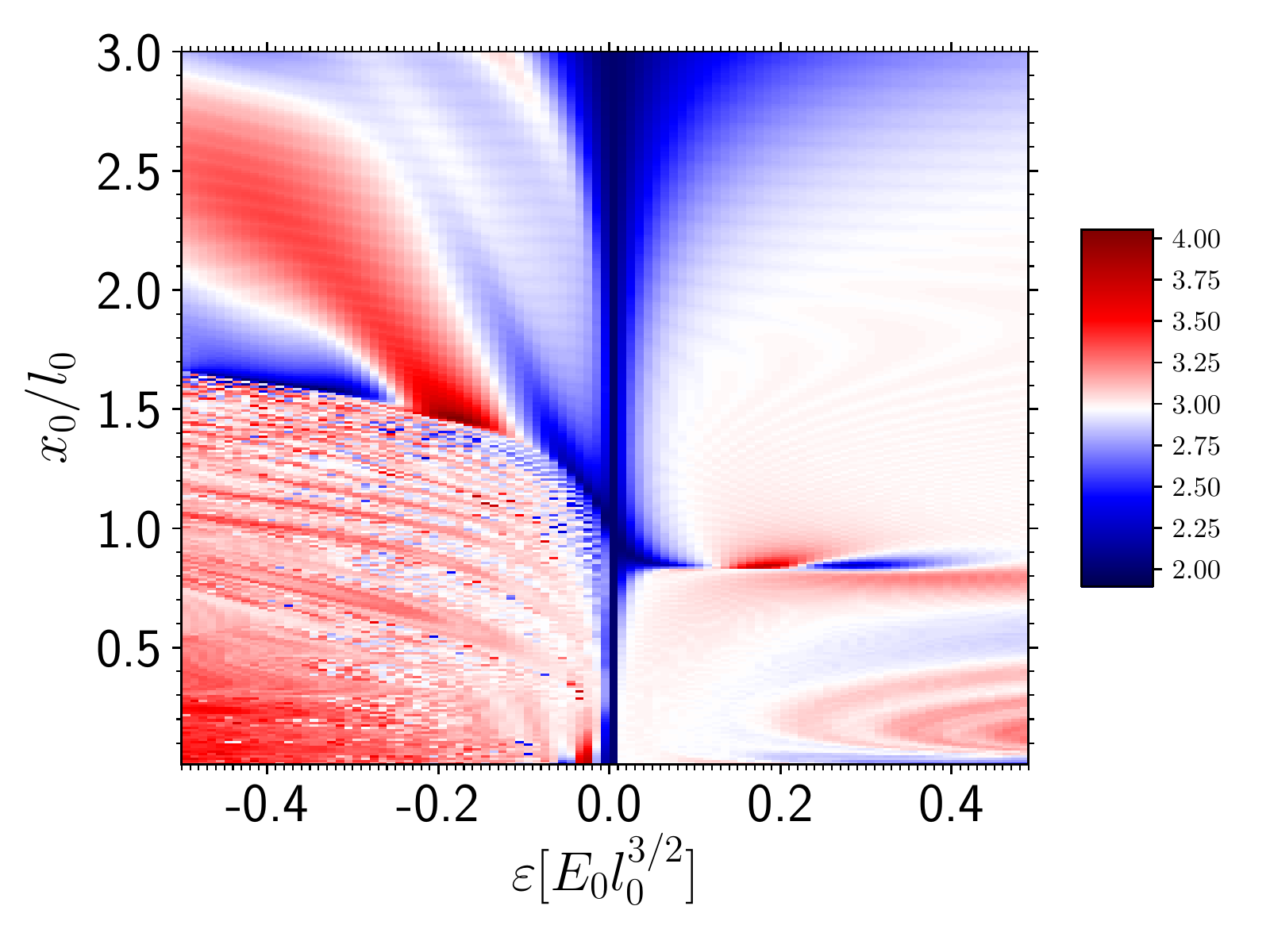}
\caption{\textit{The linear splitter}. The map of the amplitude of the
heavier soliton produced by the splitting of the original breather in
simulations of Eq. (\protect\ref{nls}) at $t=15T$, in the plane of
parameters $\protect\varepsilon $ and $x_{0}$. Black dashed lines,
separating different regions in the diagram, serve to highlight the difference between colors in the map.}
\label{lin_ampl1}
\end{figure}

The results for the potential well ($\varepsilon <0$) are summarized, along
with their counterparts for the barrier ($\varepsilon >0$) in the amplitude-
and velocity-distribution maps displayed in Figs.~\ref{lin_ampl1}-\ref%
{lin_vel2}. The maps show that the regions for $\varepsilon <0$ feature more
elaborate structure than $\varepsilon >0$, and, as mentioned above, much
smaller velocities. In fact, a larger part of the velocity maps for $%
\varepsilon <0$ is accounted for by small oscillations of the trapped
soliton with amplitude $A_{1}$.

The comparison of the analytical energy-balance prediction for the case of
the potential well, given by Eq.~\eqref{c2}, with the numerical results
demonstrates that the analytical prediction overestimates velocity $c_{2}$.
The discrepancy is explained by the fact that the radiation shed off by the
splitting breather carries away a large portion of the energy, which is not
taken into account by the analytical approximation.

\subsection{The nonlinear splitter}

\begin{figure}[th]
\centering
\includegraphics[width=0.40\textwidth]{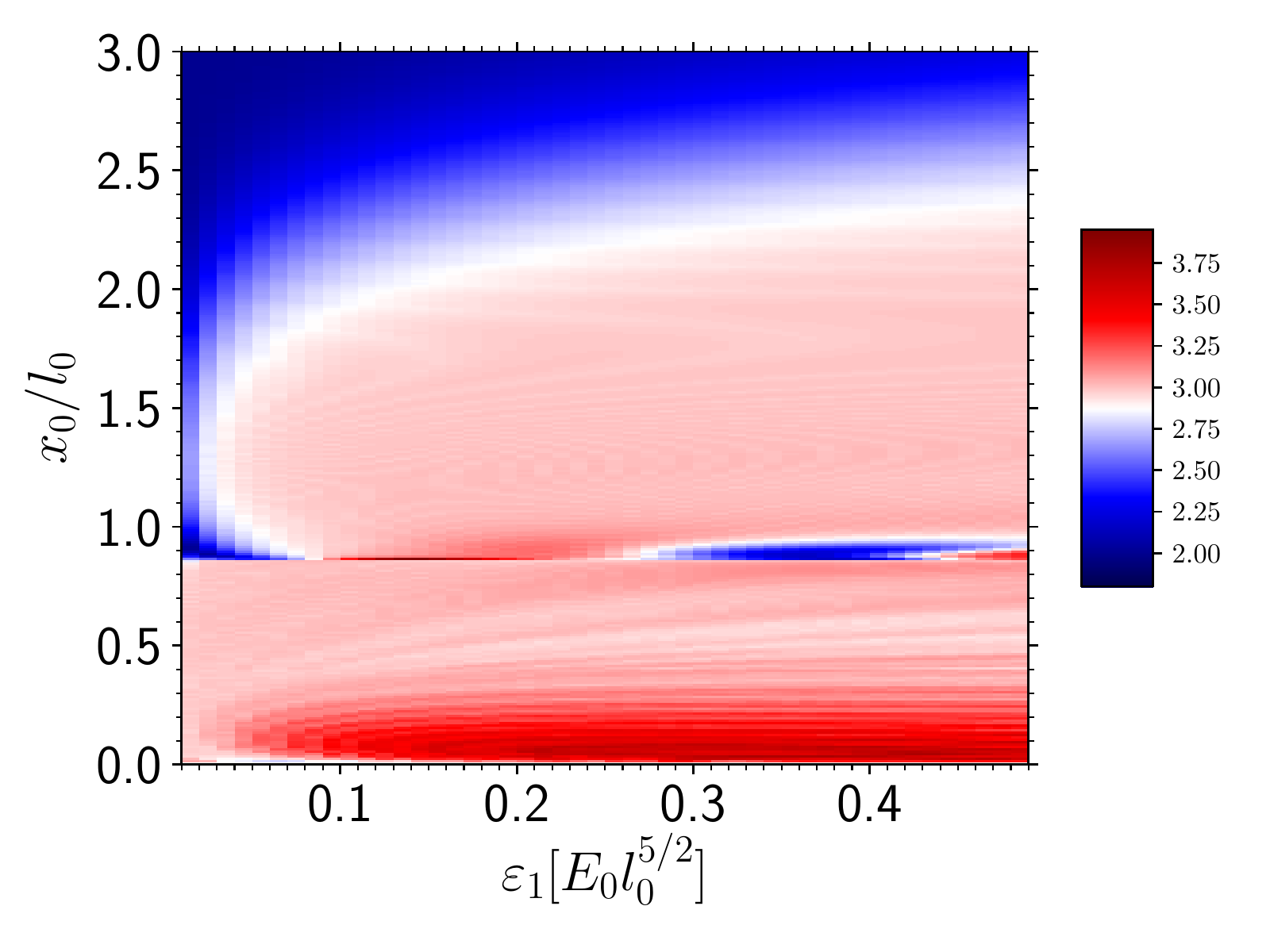}
\caption{\textit{The nonlinear splitter.} The map of the amplitude of the
larger soliton produced by the splitting of the original breather in
simulations of Eq. (\protect\ref{nonlin}) at $t=15T$, in the plane of
parameters $\protect\varepsilon _{1}$ and $x_{0}$. }
\label{nl_ampl1}
\end{figure}

\begin{figure}[th]
\centering
\includegraphics[width=0.40\textwidth]{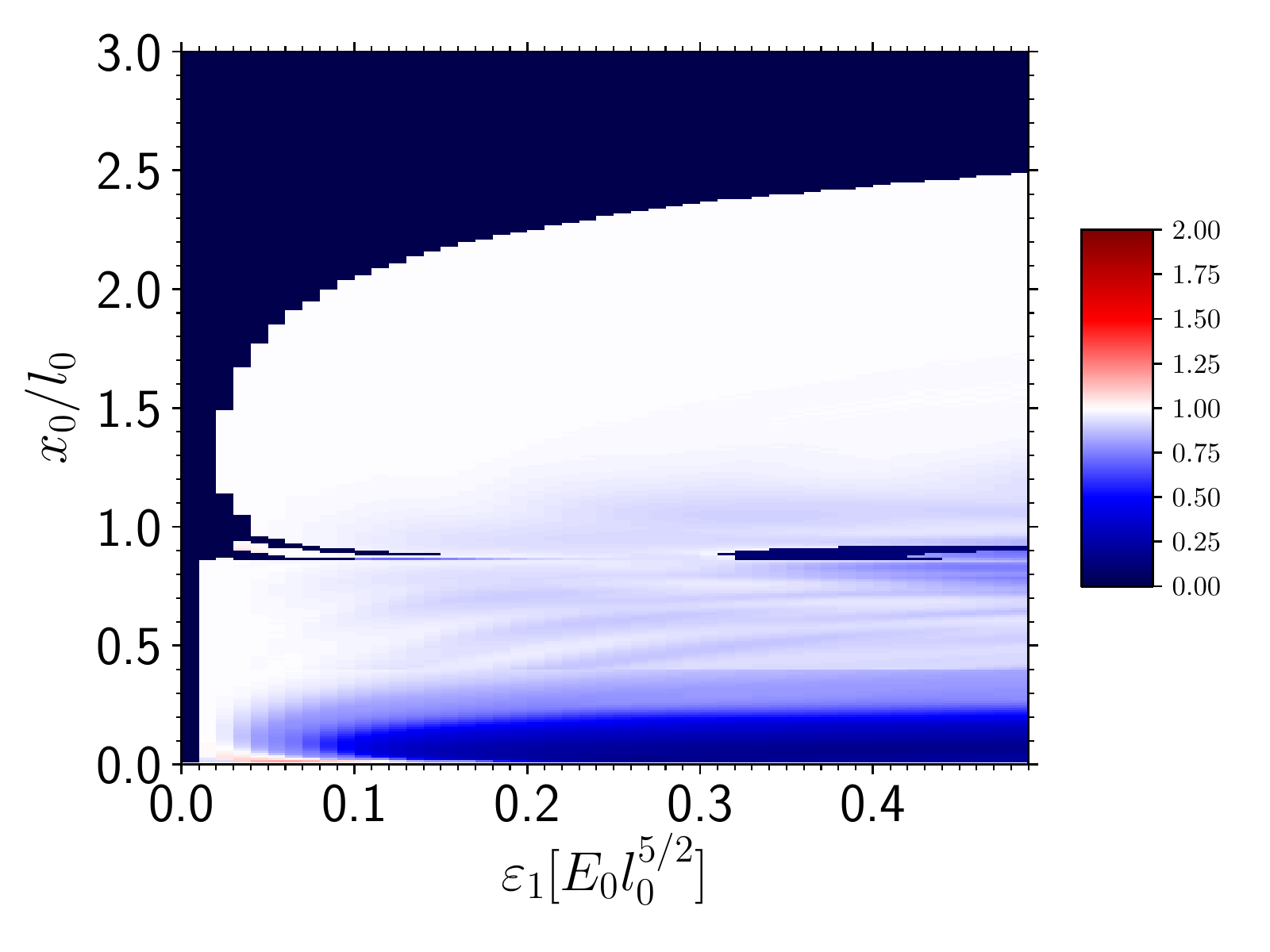}
\caption{\textit{The nonlinear barrier.} The same as in Fig. \protect\ref%
{nl_ampl1}, but for the amplitude of the lighter soliton produced by the
splitting of the original breather.}
\label{nl_ampl2}
\end{figure}

\begin{figure}[th]
\centering
\includegraphics[width=0.40\textwidth]{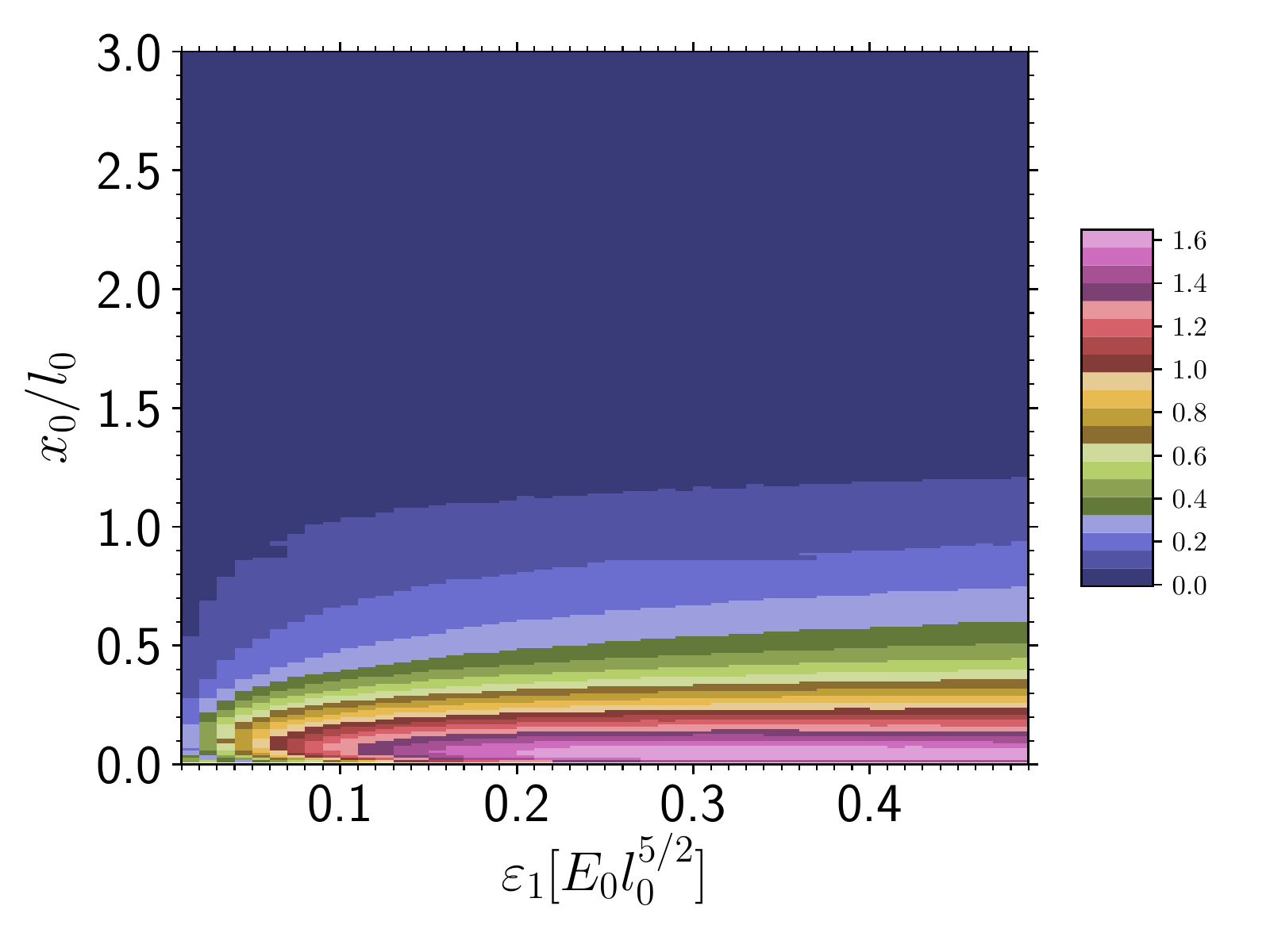}
\caption{\textit{The nonlinear splitter.} The map of values of the velocity
of the heavier soliton, produced by the splitting of the original breather
in simulations of Eq. (\protect\ref{nonlin}) at $t=15T$, in the plane of
parameters $\protect\varepsilon _{1}$ and $x_{0}$.}
\label{nl_vel1}
\end{figure}

\begin{figure}[th]
\centering
\includegraphics[width=0.40\textwidth]{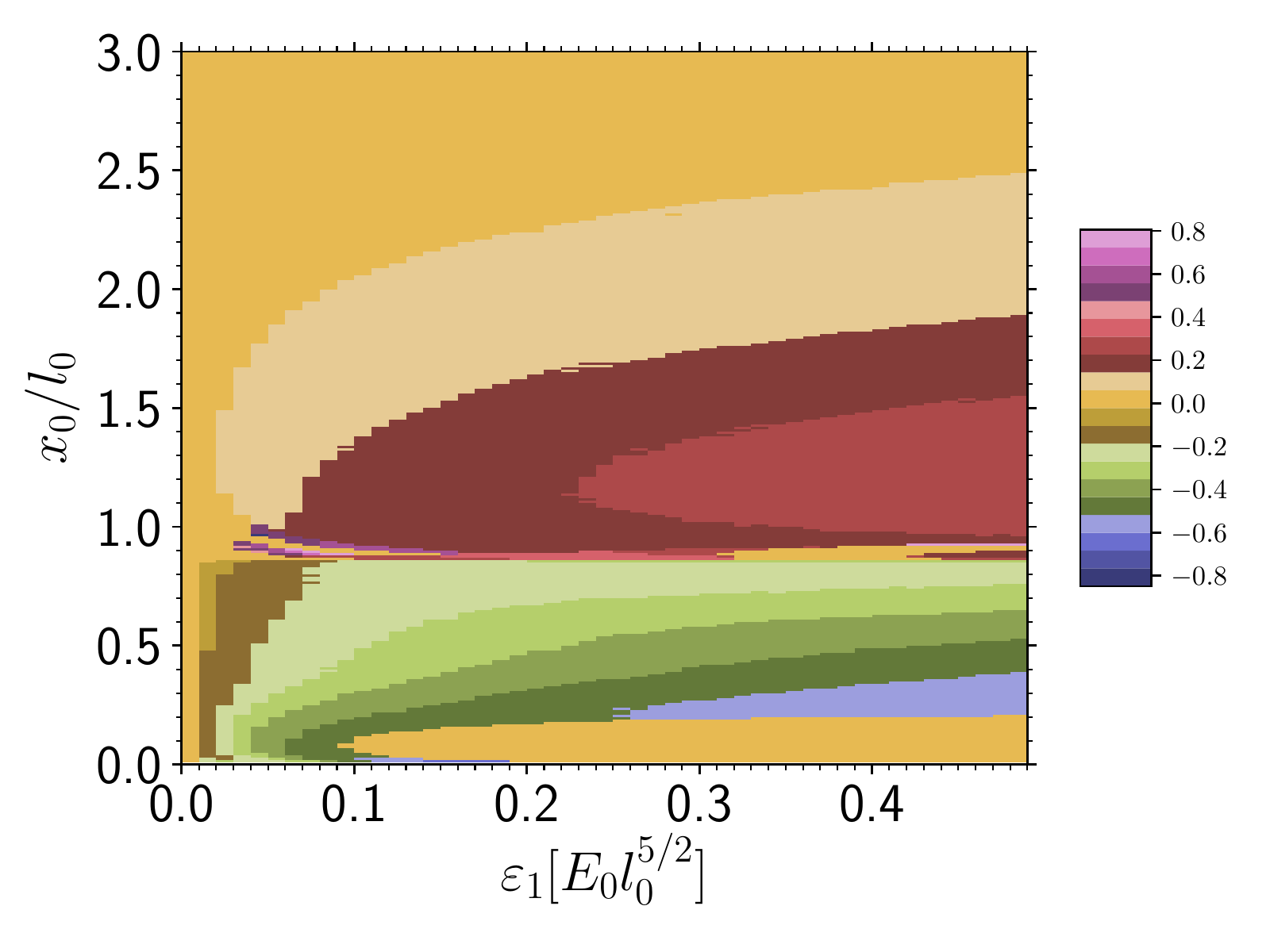}
\caption{\textit{The nonlinear barrier.} The same as in Fig. \protect\ref%
{nl_vel1}, but for the velocity of the lighter soliton (the one with the
smaller amplitude).}
\label{nl_vel2}
\end{figure}

The results are summarized in the form of maps for amplitude and velocities
distributions in the plane of $\left( \varepsilon _{1},x_{0}\right) $, which
are displayed in Figs.~\ref{nl_ampl1}-\ref{nl_vel2}. In comparison to the
results reported above for the linear barrier, the boundary between
splitting and non-splitting regimes (where the breather bounces from the
barrier and thus avoids the splitting) is much sharper in the present case,
although it remains close to approximately the same critical position, $%
x_{0}^{\mathrm{(crit)}}\approx 0.9$, see Eq. (\ref{0.9}). Another difference
is that the amplitude ratio of the fragments tends to deviate from the
\textquotedblleft natural" ratio $3:1$ (e.g., ratio $\simeq 10:1$
corresponding to $x_{0}=0$), and their velocities are much smaller than in
the case of the linear splitter.

\end{document}